\documentclass{iopart}

\usepackage{amssymb,xspace,graphicx,ifthen,amsfonts}
\usepackage[utf8]{inputenc}
\usepackage{amsfonts}
\usepackage{color}
\usepackage{colordvi}
\usepackage{pifont}
\usepackage{placeins}
\usepackage{appendix}
\usepackage{dsfont}
\usepackage[square, comma, numbers, sort&compress]{natbib}

\usepackage[colorlinks=true,citecolor=green,linkcolor=red,urlcolor=blue]{hyperref}

\begin{document}
 
\title{Breakdown of a perturbed $\mathbb{Z}_N$ topological phase}

\author{Marc Daniel Schulz}
\address{Lehrstuhl f\"ur Theoretische Physik I, Technische Universit\"at Dortmund, Otto-Hahn-Stra\ss e 4, 44221 Dortmund, Germany}
\ead{schulz@fkt.physik.tu-dortmund.de}

\author{S\'{e}bastien Dusuel}
\address{Lyc\'ee Saint-Louis, 44 Boulevard Saint-Michel, 75006 Paris, France}
\ead{sdusuel@gmail.com}

\author{Roman Or\'us}
\address{Max-Planck-Institut f\"ur Quantenoptik, Hans-Kopfermann-Stra\ss e 1, 85748 Garching, Germany}
\ead{roman.orus@mpq.mpg.de}

\author{Julien Vidal}
\address{Laboratoire de Physique Th\'eorique de la Mati\`ere Condens\'ee,
CNRS UMR 7600, Universit\'e Pierre et Marie Curie, 4 Place Jussieu, 75252
Paris Cedex 05, France}
\ead{vidal@lptmc.jussieu.fr}

\author{Kai Phillip Schmidt}
\address{Lehrstuhl f\"ur Theoretische Physik I, Technische Universit\"at Dortmund, Otto-Hahn-Stra\ss e 4, 44221 Dortmund, Germany}
\ead{schmidt@fkt.physik.tu-dortmund.de}

\begin{abstract} We study the robustness of a generalized Kitaev's toric code with $\mathbb{Z}_N$ degrees of freedom in the presence of local perturbations. For $N=2$, this model reduces to the conventional toric code in a uniform magnetic field. A quantitative analysis is performed for the perturbed $\mathbb{Z}_3$ toric code by applying a combination of high-order series expansions and variational techniques. We provide strong evidences for first- and second-order phase transitions between  topologically-ordered  and polarized phases. Most interestingly, our results also indicate the existence of topological multi-critical points in the phase diagram.
\end{abstract}


\maketitle


\section{Introduction}
The concept of topologically-ordered quantum matter  has been introduced by Wen in the context of high-temperature superconductivity \cite{Wen89,Wen90_1}, and is crucial to characterize fractional quantum Hall states and topological insulators.
 The most striking property of topologically-ordered quantum phases is their dependence on non-local properties of the system. As a consequence, such phases cannot be characterized by a local order parameter, so that the celebrated Landau's symmetry-breaking theory cannot be used.
 
More recently, topological order has attracted great interest in the field of quantum information due to its weak sensitivity to any local perturbation \cite{Kitaev03,Klich10,Bravyi10}. Indeed, non-local degrees of freedom associated with this exotic order have been shown to be (topologically) protected  against local sources of decoherence. This key idea is at the heart of  topological quantum computation \cite{Kitaev03,Ogburn99}.  
It is thus of importance to quantify precisely this protection when perturbations are added. One prominent example where the effect of additional perturbations has been extensively discussed is the case of Kitaev's toric code \cite{Kitaev03}. The toric code is an exactly solvable two-dimensional quantum spin model with a $\mathbb{Z}_2$ 
spin-liquid ground state possessing gapped (Abelian) anyonic excitations. It can be considered as one of the simplest models displaying topological order. Apart from the influence of temperature \cite{Castelnovo07,Nussinov08,Alicki09,Iblisdir09,Iblisdir10} and of disorder \cite{Wootton11,Stark11},  several works have investigated the effect of an external magnetic field in this model \cite{Trebst07,Hamma08,Tupitsyn10,Vidal09_1,Vidal09_2,Dusuel11,Tagliacozzo11}. Interestingly, a very rich phase diagram containing first- and second-order phase transitions, multi-criticality, self-duality, and dimensional reduction has been found. Similar phase transitions out of topologically-ordered phases have been studied in the context of the Levin-Wen model \cite{Levin05,Burnell12,Burnell11}.

Actually, the toric code model can be defined for any discrete Abelian or non-Abelian group \cite{Kitaev03,Bais09,Wootton11_1}. In this work, we present an extension of this model to  $\mathbb{Z}_N$ degrees of freedom \cite{Bullock07}, which reduces to the conventional toric code for $N=2$. Although excitations are still Abelian, qualitative and quantitative analyses of this model in the presence of local perturbations provide some insights in the understanding of topological phase transitions when more complex degrees of freedom are involved.

The paper is organized as follows~: in \sref{sec:ZN}, we describe an extension of Kitaev's toric code from $\mathbb{Z}_2$ to $\mathbb{Z}_N$ degrees of freedom. Our starting point is Wen's plaquette model \cite{Wen03} for which a generalization to $\mathbb{Z}_N$ can be written down rather easily. The $\mathbb{Z}_N$ plaquette model is then mapped onto a $\mathbb{Z}_N$ toric code which is shown to display  topological order and  anyonic statistics.  

To study the robustness of the topological order,  we analyze the influence of local perturbations in \sref{sec:ZNF}. For $N=2$, such perturbations correspond to a uniform magnetic field. If the perturbations are strong enough, one expects conventional polarized phases that are not topologically ordered. As a consequence, a  phase transition between the topological and the polarized phases must occur. Analyzing the breakdown of the topological phase is a very challenging problem for general~$N$. Nevertheless, as we shall see for special kinds of perturbation, either exact mappings onto already known models exist or the model displays self-duality and dimensional reduction.

In \sref{sec:Z3F}, we focus on the case $N=3$ and probe the robustness of the  $\mathbb{Z}_3$ topological phase for simple perturbations. To this end, we use perturbative continuous unitary transformations (pCUT) and a variational approach based on infinite projected entangled pair states (iPEPS). Note that this combined pCUT+iPEPS method has been already used successfully for the standard toric code $(N=2)$ in an arbitrary magnetic field \cite{Dusuel11}. After a brief discussion of both methods and their combination, results are presented for two types of perturbation that can be viewed as natural generalizations of the $N=2$ toric code either in a parallel field or in a transverse field.
In the simplest case where the perturbation commutes with local charge (or flux) operators, we establish an exact mapping, valid at low-energies, onto a three-state clock model in a transverse field. For a ferromagnetic coupling, this model is known to display a weakly first-order transition (see for instance \cite{Hamer92}). However, the perturbation considered here also leads us to study the antiferromagnetic three-state clock model in a transverse field which, to our knowledge,  has never been discussed in the literature. In this work, we find evidence for a second-order transition in this system (whose universality class still remains to be accurately determined).
We then discuss the counterpart of an arbitrary parallel magnetic field in the $N=2$ model. For such a perturbation, we obtain a rich phase diagram containing first- and second-order phase transition lines that form the boundary of the topological phase. Finally, we study the transverse-field problem that diplays dimensional reduction and self-duality, as the $N=2$ model \cite{Vidal09_2}. Conclusions and perspectives are drawn in \sref{sec:Conclusion}.

\section{Construction of the generalized $\mathbb{Z}_N$ toric code}\label{sec:ZN}
In the following, we first introduce the $\mathbb{Z}_N$-generalization of the plaquette model introduced by Wen \cite{Wen03}. The main reason to do so is that the plaquette model with $\mathbb{Z}_N$ degrees of freedom arises rather naturally from the $\mathbb{Z}_2$ conventional counterpart and it is very simple to write down. Afterwards we perform a mapping to a generalized toric code model  with $\mathbb{Z}_N$-Abelian anyons \cite{Bullock07}.
%
\subsection{$\mathbb{Z}_N$ plaquette model}
\begin{figure}[ht]
\centering
  \includegraphics[width=10cm]{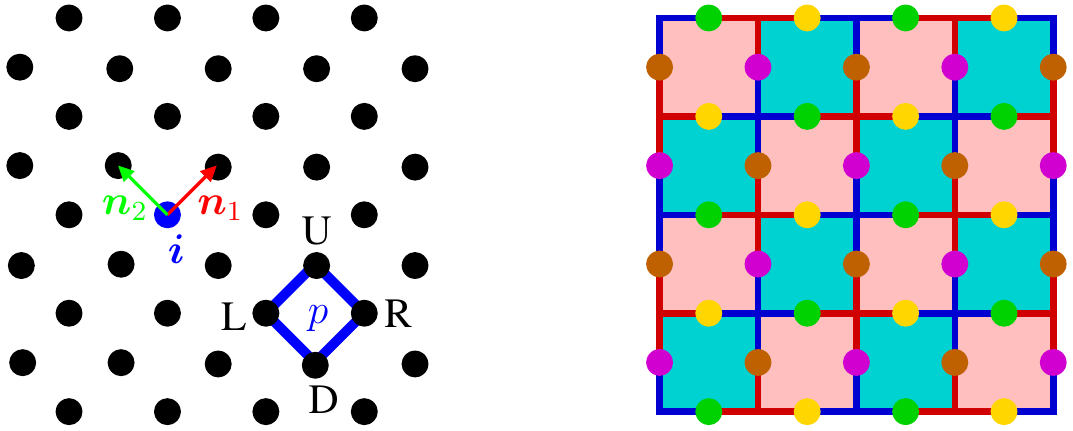} 
  \caption{Left~: a piece of the square lattice on which one defines Wen's plaquette model. Right~: a piece of the four-colored square lattice on which one defines Kitaev's toric code model with $\mathbb{Z}_N$ degrees of freedom. In Wen's model, degrees of freedom are located on the vertices of the lattice whereas in Kitaev's model, they live on the bonds of the lattice.}
  \label{fig:square_lattice}
\end{figure}

%
We consider $N$-state degrees of freedom located on the sites $\bi{i}$ of a square lattice whose unit-cell vectors $\bi{n}_1$ and $\bi{n}_2$ are shown in \fref{fig:square_lattice} (left). The associated orthonormal states are denoted by $|q\rangle_\bi{i}$, where $q\in\mathbb{Z}_N$.
Next, let us  define the operators $Z_\bi{i}$ and $X_\bi{i}$ as 
\begin{equation}
\label{eq:Definition_ZX_Operators}
 	Z_\bi{i} |q\rangle_\bi{i} = \omega^q |q\rangle_\bi{i} \quad \textrm{and} \quad X_\bi{i} |q\rangle_\bi{i} = |q-1 \rangle_\bi{i} \, ,
\end{equation}
where $\omega=\rme^{\frac{2\mathrm{i} \pi }{N}}$. These unitary operators reduce to the conventional Pauli matrices 
$\sigma^{z}_\bi{i}$ and $\sigma^{x}_\bi{i}$ for $N=2$. On the same site, both operators obey the important ``commutation relation'' (Weyl algebra) 
\begin{equation}\label{eq:Elementary_Commutation_Relation}
X_\bi{i} Z_\bi{i} =\omega Z_\bi{i} X_\bi{i} \, ,
\end{equation}
which generalizes the well-known anticommutation relation $\sigma^{x}_\bi{i}\sigma^{z}_\bi{i} = - \sigma^{z}_\bi{i}\sigma^{x}_\bi{i}$ of Pauli matrices. They obviously commute when acting on different sites.

The Hamiltonian of the $\mathbb{Z}_N$ plaquette model is then defined by
\begin{equation}\label{eq:Plaquette_Model}
 	H_{\rm{plaquette}}=-J \sum_{p} \left( W_p^{\phantom{\dagger}} + W_p^{\dagger} \right) \, ,  
\end{equation}
where $W_p= Z_{\mathrm{D}} X_{\mathrm{R}} Z_{\mathrm{U}} X_{\mathrm{L}}$. Sites D, R, U, and L correspond to the four sites $(\bi{i},\bi{i}+\bi{n}_1,\bi{i}+\bi{n}_1+\bi{n}_2,\bi{i}+\bi{n}_2)$ of an elementary plaquette $p$ of the square lattice (see \fref{fig:square_lattice}).  

Using (\ref{eq:Definition_ZX_Operators}) and (\ref{eq:Elementary_Commutation_Relation}), it is easy to check that 
\begin{equation}
\label{eq:Wp_cons}
 	\left[ W^{\phantom{\dagger}}_{p^{\phantom{\prime}}}, W^{\phantom{\dagger}}_{p^{\prime}}\right]=\left[ W^{\phantom{\dagger}}_{p^{\phantom{\prime}}}, W^{\dagger}_{p^{\prime}}\right] = 0 \, , 
\end{equation}
for all $p$ and $p'$, so that all plaquette operators commute with the Hamiltonian. Furthermore, it is important to note that these operators obey $W_p^N=\mathds{1}$. In other words, $W_p$'s are $\mathbb{Z}_N$ conserved quantities whose eigenvalues are simply 
$\{ \omega^q,q\in \mathbb{Z}_N \}$. As for the standard $\mathbb{Z}_2$ plaquette model \cite{Wen03}, this property ensures  exact solvability of $H_{\rm{plaquette}}$. In the following, we map the plaquette model onto a generalized toric code \cite{Kitaev03,Bullock07} as was already done for the $\mathbb{Z}_2$ case \cite{Kitaev06,Vidal08_2}. The main advantage of the toric code is that the quantum statistics of the elementary excitations are simpler to identify. Additionally, it allows one to adopt an {\it ad hoc} language reminiscent of lattice gauge theories with $\mathbb{Z}_N$ degrees of freedom (see, {\it e.g.}, \cite{Arakawa04,Doucot04}).

%
\subsection{$\mathbb{Z}_N$ toric code}
%
\subsubsection{Mapping}
In order to map model (\ref{eq:Plaquette_Model}) onto a generalized toric code model \cite{Kitaev03}, we introduce a translationally invariant and four-colored lattice as depicted on the right side of \fref{fig:square_lattice}, where the $N$-state degrees of freedom of the plaquette model are placed on the bonds of a square lattice. We then perform the following local, unitary transformations~:
\begin{eqnarray}
  	X^{\phantom{\dagger}}_{\bi{i} \in \mathrm{brown, magenta}} \rightarrow  Z^{\phantom{\dagger}}_{\bi{i}} \, , & \quad & X^{\phantom{\dagger}}_{\bi{i} \in  \mathrm{yellow, green}} \rightarrow  X^{\phantom{\dagger}}_{\bi{i}} \, , \nonumber\\
  	Z^{\phantom{\dagger}}_{\bi{i} \in \mathrm{brown, magenta}} \rightarrow  X^{\dagger}_{\bi{i}} \, , &\quad & Z^{\phantom{\dagger}}_{\bi{i} \in \mathrm{yellow, green}} \rightarrow  Z^{\phantom{\dagger}}_{\bi{i}}  \, .
\end{eqnarray}
As a consequence, the plaquette operators $W_p$ become different on the red and blue stars $s$ and on the cyan and pink plaquettes $p$ as illustrated in \fref{fig:square_lattice}. To strengthen the analogy with the conventional toric code, we relabel the various operators as follows~: 
\begin{eqnarray}
    \label{eq:mapping_to_AB}
  	\fl \quad W^\dagger_{p \in \mathrm{red{\phantom{s}}}} \rightarrow A_{s \in \mathrm{red{\phantom{s}}}} = X^{\phantom{\dagger}}_{\mathrm{D}} X^{{\dagger}}_{\mathrm{R}} X^{\phantom{\dagger}}_{\mathrm{U}} X^{{\dagger}}_{\mathrm{L}}  \, , & \quad & W^{\phantom{\dagger}}_{p \in \mathrm{cyan}} \rightarrow B_{p \in \mathrm{cyan}} = Z^{\phantom{\dagger}}_{\mathrm{D}} Z^{\phantom{\dagger}}_{\mathrm{R}} Z^{\phantom{\dagger}}_{\mathrm{U}} Z^{\phantom{\dagger}}_{\mathrm{L}}  \, ,  \nonumber\\
  	\fl \quad  W^{\phantom{\dagger}}_{p \in \mathrm{blue}} \rightarrow A_{s \in \mathrm{blue}} = X^{{\dagger}}_{\mathrm{D}} X^{\phantom{\dagger}}_{\mathrm{R}} X^{{\dagger}}_{\mathrm{U}} X^{\phantom{\dagger}}_{\mathrm{L}}  \, , & \quad & W^\dagger_{p \in \mathrm{pink}} \rightarrow B_{p \in \mathrm{pink}} = Z^{{\dagger}}_{\mathrm{D}} Z^{\dagger}_{\mathrm{R}} Z^{\dagger}_{\mathrm{U}} Z^{\dagger}_{\mathrm{L}}  \, .
\end{eqnarray}
Let us underline that the four-coloring of the lattice is mandatory if one wants to define $B_p$ with $Z$ (or $Z^{{\dagger}}$) operators only. Nevertheless, other choices with smaller units cells are possible.

Finally, we obtain the Hamiltonian of the $\mathbb{Z}_N$ toric code
\begin{eqnarray}\label{eq:TC_Hamilton}
 	H_{\rm{TC}}=-J \sum_{s} \left( A_s^{\phantom{\dagger}} + A_s^{\dagger} \right) -J \sum_{p} \left( B_p^{\phantom{\dagger}} + B_p^{\dagger} \right) \, .
\end{eqnarray}

Let us note that this model was already introduced in \cite{Bullock07} but differs from
the $\mathbb{Z}_N$ toric code discussed in \cite{Kitaev03} which involves projectors $\mathcal{P}_s$ and 
$\mathcal{P}_p$ instead of $A_s+A_s^\dagger$ and $B_p+B_p^\dagger$ (see next section for definitions). However, since both models are equivalent for $N=2,3$, we shall (abusively) call them toric codes.
%
\subsubsection{Ground states and topological degeneracy}
\label{sec:gs}
%

A direct consequence of (\ref{eq:Wp_cons}) is that all $A_s$ and
$B_{p}$ operators commute with $H_{\rm{TC}}$. Thus, the ground-state energy (per site) $e_0$ is simply obtained by choosing the (possibly degenerate) minimal eigenvalue of the local operators $-J(A_s+A_s^\dagger)$ or $-J(B_p+B_p^\dagger)$, namely, $e_0=-2J\cos(2 \pi k/N)$. For $J>0$ and for any $N$, the ground state is  unique and obtained for $k=0$. However, for $J<0$, the ground state is unique for $N$ even (in this case, one chooses $k=N/2$) but it is infinitely-many degenerate for $N$ odd since, locally, one can choose $k=(N\pm 1)/2$. In the following, we will only consider the simplest case $J>0$.

Nevertheless, there are subtleties  since the  ground-state degeneracy also depends on the surface's topology as we shall now see on two simple examples.
Let us first consider an infinite open plane for which no constraint on $A_s$'s and $B_{p}$'s exists. In this case, the ground state is unique and can be  built as
\begin{equation}
\label{eq:TC_gs_OBC}
 	\left| \mathrm{gs} \right\rangle = \mathcal{N} \prod_s \mathcal{P}_s \prod_p \mathcal{P}_p \left| \mathrm{ref}\right\rangle \, , 
\end{equation}
where $\mathcal{N}$ is a normalization constant and 
%
%
\begin{equation}
\mathcal{P}_s=\frac{1}{N}\sum_{k =0}^{N-1} A_s^k  \ , \
\mathcal{P}_p=\frac{1}{N}\sum_{k =0}^{N-1} B_p^k \, .
\end{equation}
%
%
Operators $\mathcal{P}_s$ ($\mathcal{P}_p$) project on subspaces with eigenvalue $1$ of the corresponding $A_s$ ($B_p$). The reference state $\left|\mathrm{ref}\right\rangle$ can be chosen arbitrarily provided it leads to $ \left| \mathrm{gs} \right\rangle \neq 0$. For instance, one may choose the fully-polarized state
$\displaystyle{\left|\mathrm{ref}\right\rangle = \bigotimes_{\bi{i}} \ \left|0\right\rangle_{\bi{i}}}$ that already fulfills 
$\mathcal{P}_p\left|\mathrm{ref}\right\rangle=\left|\mathrm{ref}\right\rangle$ for all plaquettes $p$.

\begin{figure}[ht]
\centering
  \includegraphics[width=10cm]{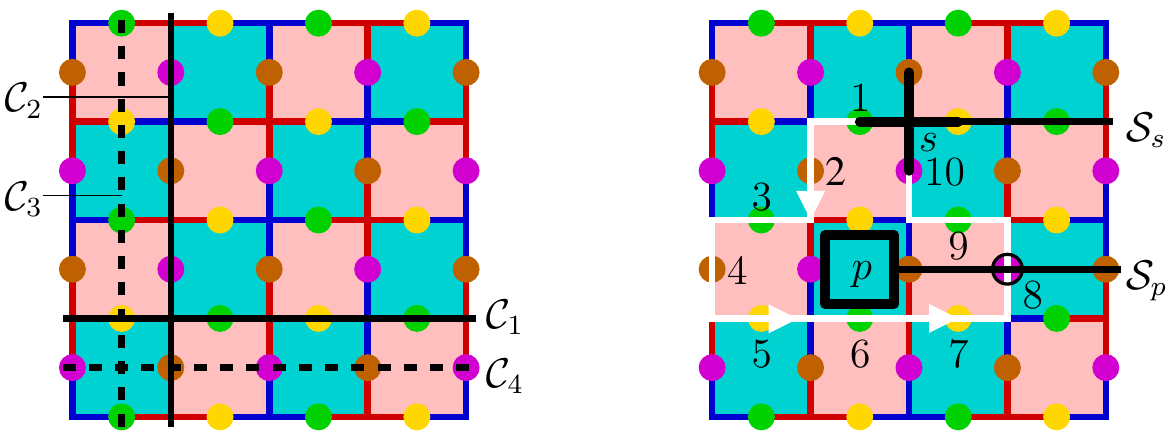} 
  \caption{Left~: possible choices of non-contractible loops $\mathcal{C}_k$
  with $k\in\{1,2,3,4\}$ for a system with periodic boundary conditions.
  Right~: illustration of the semi-infinite strings $\mathcal{S}_s$ and
  $\mathcal{S}_p$ used to define charge and flux states in
  (\ref{eq:one_particle_states}) and (\ref{eq:one_particle_states_2}).
  An example of a counter-clockwise braiding contour for moving a charge $q_s$
  initially located on a red star $s$ around a flux $q_p$ located on a cyan
  plaquette $p$ is represented in white (see text for explanations).
  The black circle locates the site (numbered 8) where the crossing between the
  braiding contour and the string $\mathcal{S}_p$ of the flux occurs.
}
  \label{fig:cycles}
\end{figure}

Next, let us consider the $\mathbb{Z}_N$ toric code on a torus. In this case, there are two  constraints 
\begin{eqnarray}\label{eq:constrained}
 	\prod_s A_s = \mathds{1} \, , \quad \prod_p B_p=\mathds{1} \, ,
\end{eqnarray}
so that the number of independent eigenvalues of $A_s$ and $B_{p}$ operators is reduced by two. 
However, as in the $\mathbb{Z}_2$ toric code, there exist conserved loop operators that can be chosen as~:
\begin{eqnarray} 
	\fl
	\mathcal{Z}^{\phantom{\dagger}}_1=
	\left(\prod_{\bi{i}\in\mathcal{C}_{1},\mathrm{green}}\!\!Z^{\phantom{\dagger}}_\bi{i}\right)\!
	\left(\prod_{\bi{i}\in\mathcal{C}_{1},\mathrm{yellow}}\!\!Z_\bi{i}^\dagger\right)\, , &
	\mathcal{Z}^{\phantom{\dagger}}_2=\left(\prod_{\bi{i}\in\mathcal{C}_{2},\mathrm{magenta}}\!\!Z^{\phantom{\dagger}}_\bi{i}\right)\!
	\left(\prod_{\bi{i}\in\mathcal{C}_{2},\mathrm{brown}}\!\!Z_\bi{i}^\dagger\right), \nonumber\\
	\fl
	\mathcal{X}^{\phantom{\dagger}}_1=
	\left(\prod_{\bi{i}\in\mathcal{C}_{3},\mathrm{green}}\!\!X^{\phantom{\dagger}}_\bi{i}\right)\!
	\left(\prod_{\bi{i}\in\mathcal{C}_{3},\mathrm{yellow}}\!\!X_\bi{i}^\dagger\right) , \ &
	\mathcal{X}^{\phantom{\dagger}}_2=
	\left(\prod_{\bi{i}\in\mathcal{C}_{4},\mathrm{magenta}}\!\!X^{\phantom{\dagger}}_\bi{i}\right)\!
	\left(\prod_{\bi{i}\in\mathcal{C}_{4},\mathrm{brown}}\!\!X_\bi{i}^\dagger\right)\!,
\end{eqnarray}
where $\mathcal{C}_k$ with  $k\in\{1,2,3,4\}$ are the non-contractible loops
of the torus depicted in \fref{fig:cycles} (left). All these operators commute with $H_{\rm TC}$, but only two of them can be chosen independently since $\left[\mathcal{Z}_\mu,\mathcal{X}_\mu\right] \neq 0$ with $\mu\in\{1,2\}$. These two additional conserved quantities maintain the exact solvability of $H_{\rm TC}$. 
Concretely, if we choose to label states with the eigenvalues  $z_\mu=\{\omega^q\}$  ($q\in \mathbb{Z}_N$) of the operators $\mathcal{Z}_\mu$, we find that there exist $N^2$ ground states $ \left| \mathrm{gs}, z_1, z_2 \right\rangle$ that can be written as
\begin{eqnarray}
  	\left| \mathrm{gs}, \omega^{q_1}, \omega^{q_2} \right\rangle = \mathcal{N'} \prod_s \mathcal{P}_s  \left(\mathcal{X}_1^\dagger\right)^{q_1}\left(\mathcal{X}_2^\dagger\right)^{q_2} \bigotimes_{\bi{i}} \ \left|0\right\rangle_{\bi{i}} \, ,
\end{eqnarray}
where $\mathcal{N'}$ is a normalization constant. 
More generally, following \cite{Kitaev03,Bullock07}, one can show that for a compact surface with genus $g$, each eigenstate is $N^{2g}$-degenerate (at least) so that the system is indeed topologically ordered.

%
\subsubsection{Excitations and statistics}
\label{sec:sub:sub:excitations_and_statistics}
%
Excitations of the toric code correspond to states which violate the condition
that all eigenvalues of the $A_s$ or the $B_p$ operators are equal to $1$.
In other words, an elementary particle is a charge $q$ on star $s$ (a flux
$q$ on plaquette $p$) corresponding to an eigenvalue $\omega^q$, with
$q\in\mathbb{Z}_N$ and $q\neq 0$, of $A_s$ ($B_p$) (remember that $q=0$ defines the ground state).
As $A_s$ and $B_p$ operators commute with $H_{\rm{TC}}$, charges and fluxes are static excitations. Moreover, the form of the Hamiltonian implies that the energy of a many-particle state is simply the sum of the single-particle energies. In other words, charges and fluxes do not interact.
%
%
\begin{figure}[ht]
\centering
  \includegraphics[width=8cm]{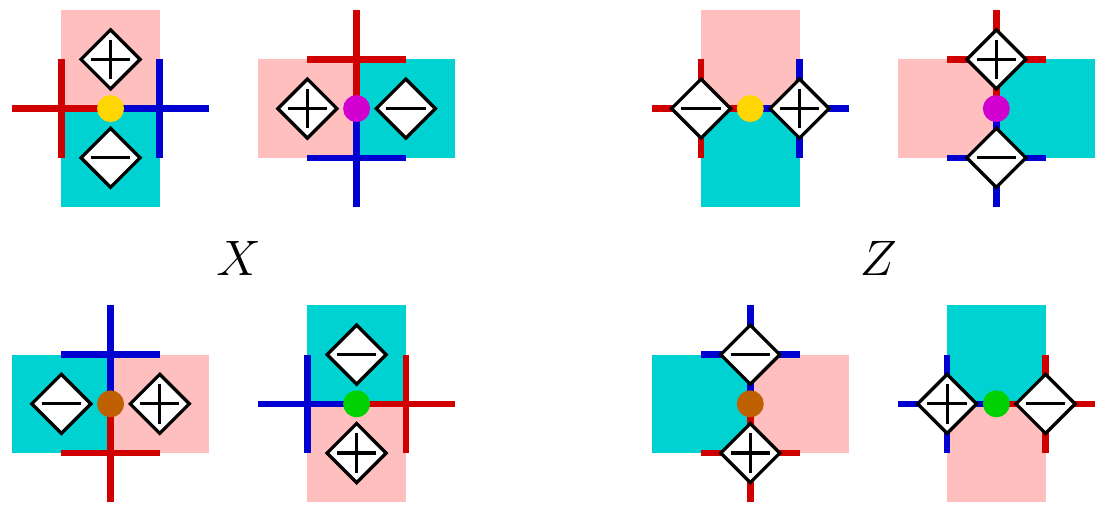} 
\caption{Illustration of the action of operators $X_\bi{i}$ (left) and
$Z_\bi{i}$ (right) on an eigenstate of $A_s$ and $B_p$ operators. A
diamond on a star or a plaquette, with a $\pm$ sign denotes a
multiplicative change by $\omega^{\pm 1}$ of the corresponding
eigenvalue. The behavior depends on the site's color.}
\label{fig:action_simple_operators}
\end{figure}
%
%
In what follows, we give the explicit construction of single-particle states
for general $N$ following the detailed construction given in \cite{Vidal08_2} for
$N=2$.
As was already the case in the $\mathbb{Z}_2$ toric code, there is no local
operator creating a single excitation. This is illustrated in
\fref{fig:action_simple_operators} where one can see that an $X_\bi{i}$
($Z_\bi{i}$) operator creates \textit{two} excitations on neighboring plaquettes
(stars).

However, for an infinite plane with open boundary conditions, it is
possible to consider single-particle states. Indeed, in such a system, a single
excitation can be  obtained by first creating a pair of charges (fluxes) and by
taking one of the particles to infinity, at least in
principle.  It is clear that this is rather a {\it gedankenexperiment} that
cannot be implemented practically (for instance, in a computer).
 Nevertheless,
one may always consider a state where the two particles originating from the
elementary pair-creation process are so distant that they are eventually
independent. Thus, a one-particle state $\left| q\right\rangle_\alpha$ with charge (flux) $q$ on $\alpha=s$ ($p$) can be defined, for instance, as 
\begin{eqnarray}
    \label{eq:one_particle_states}
	\quad \left| q\right\rangle_s =&
  	\left(\prod_{\bi{i}\in\mathcal{S}_{s},\mathrm{green}}\!\!Z_\bi{i}\right)\!
  	\left(\prod_{\bi{i}\in\mathcal{S}_{s},\mathrm{yellow}}\!\!Z_\bi{i}^\dagger\right)
    \left| \mathrm{gs} \right\rangle, \\
    \label{eq:one_particle_states_2}
	\quad \left| q\right\rangle_p =&
    \left(\prod_{\bi{i}\in\mathcal{S}_{p},\mathrm{magenta}}\!\!X_\bi{i}\right)\!
  	\left(\prod_{\bi{i}\in\mathcal{S}_{p},\mathrm{brown}}\!\!X_\bi{i}^\dagger\right)
    \left| \mathrm{gs} \right\rangle
\end{eqnarray}
where the semi-infinite strings $\mathcal{S}_\alpha$ are displayed in \fref{fig:cycles} (right).

We shall now show on a specific but sufficiently general  example that 
charges and fluxes obey mutual anyonic statistics. Let us consider an eigenstate, denoted by
$\left| \psi \right\rangle$, with a charge $q_s$ and a flux $q_p$ at positions
shown in \fref{fig:cycles} (right). One can braid the
charge around the flux along the counter-clockwise oriented  white path drawn in
this figure, by acting on $\left| \psi \right\rangle$ with the operator
$\mathcal{O}=Z^{q_s}_{10} \ldots Z^{q_s}_{7} {\left(Z^{q_s}_6\right)}^\dagger
Z^{q_s}_{5} \ldots Z^{q_s}_{1}$.
This can be checked from the action of $Z_\bi{i}$ operators shown in
\fref{fig:action_simple_operators} (right).
Furthermore, from the definition of $B_p$ operators given in
(\ref{eq:mapping_to_AB}), this operator $\mathcal{O}$ is nothing but the
product of all $\left(B_p^\dagger\right)^{q_s}$ operators for all plaquettes
encircled by the braiding contour.
Given that $\left| \psi \right\rangle$ is an eigenstate of
all plaquette operators with eigenvalue $1$, except for the plaquette $p$ where
the flux is located and for which
$B_p\left| \psi \right\rangle = \omega^{q_p}\left| \psi \right\rangle$,
one gets
$\mathcal{O}\left| \psi \right\rangle=
\omega^{-q_s q_p}\left| \psi \right\rangle$. This non-trivial braiding phase is
the signature of the mutual ($\mathbb{Z}_N$) anyonic statistics between
charges and fluxes.
It is similar to an Aharonov-Bohm phase, which explains the terminology of
charges and fluxes employed to describe the excitations. The same argument
allows one to show that a braiding of a charge (flux) around a charge (flux) leads to a
trivial phase which is reminiscent from the bosonic statistics of charges (fluxes). In addition, it is clear that there is a hard-core constraint since one cannot create two particles on the same star or on the same plaquette. 

Let us end this discussion by
showing that the phase can be obtained in another, complementary way.
One can write explicitely the state $\left| \psi \right\rangle$ as
%
\begin{equation}
\fl
\left| \psi \right\rangle = 
\left(\prod_{\bi{i}\in\mathcal{S}_{s},\mathrm{green}}\!\!Z_\bi{i}\right)\!
\left(\prod_{\bi{i}\in\mathcal{S}_{s},\mathrm{yellow}}\!\!Z_\bi{i}^\dagger\right)
\left(\prod_{\bi{i}\in\mathcal{S}_{p},\mathrm{magenta}}\!\!X_\bi{i}\right)\!
\left(\prod_{\bi{i}\in\mathcal{S}_{p},\mathrm{brown}}\!\!X_\bi{i}^\dagger\right)
\left| \mathrm{gs} \right\rangle.
\end{equation}
%
Then, one can compute $\mathcal{O}\left| \psi \right\rangle$ by commuting
$\mathcal{O}$ with the operators appearing in the above expression using
(\ref{eq:Elementary_Commutation_Relation}) and 
$\mathcal{O}\left| \mathrm{gs} \right\rangle=\left| \mathrm{gs} \right\rangle$
since the ground-state is flux-free. For the particular braiding
represented in \fref{fig:cycles} (right), the non-trivial phase will appear from the
commutation $Z_8^{q_s} X_8^{q_p} = \omega^{-q_s q_p} X_8^{q_p} Z_8^{q_s}$
at site number 8 where the braiding contour and the string of the flux ${S}_{p}$ intersect.

\section{\texorpdfstring{$\mathbb{Z}_N$}{Z(N)} toric code in the presence of local uniform perturbations}\label{sec:ZNF}
In this section, we first define the general local perturbations for the $\mathbb{Z}_N$ toric code model. 
The presence of such perturbations destroys the exact solvability of the toric code model. As mentioned in the introduction, a phase transition has to take place when the perturbation strength increases since, for $J=0$, the ground state is fully polarized and thus not topologically ordered. 
Thereafter, we  focus on special examples that allow for a  detailed investigation of this transition for general $N$.

%
\subsection{General structure of the perturbation}
\label{sec:sub:gen_struct_pert}
%

\begin{figure}[ht]
\centering
  \includegraphics[width=4cm]{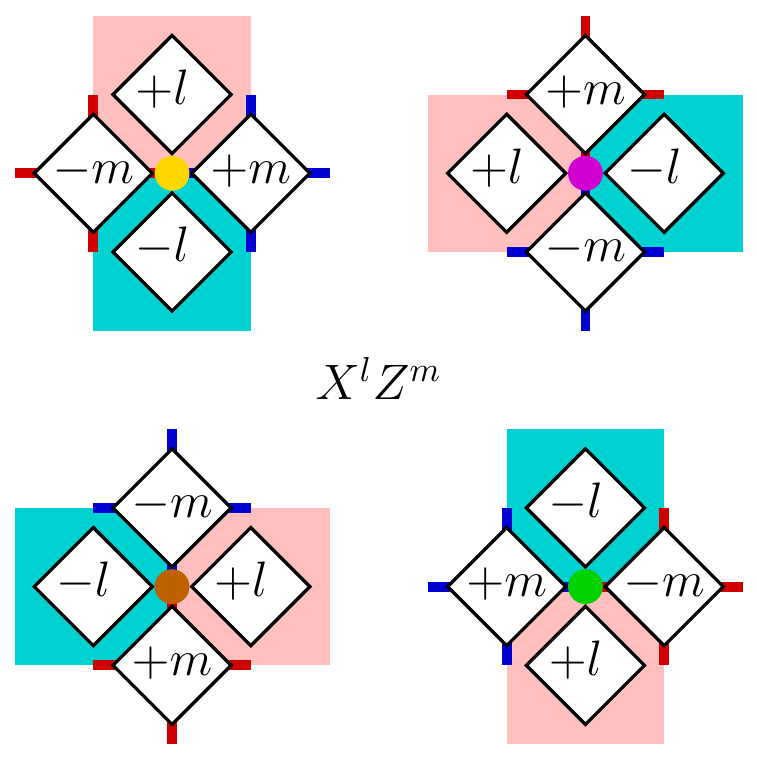} 
\caption{Illustration of the action of the operator
$X_\bi{i}^lZ_\bi{i}^m$ on an eigenstate of $A_s$ and $B_p$
operators. A diamond on a star or a plaquette, with a value $k=\pm l,\pm m$,
denotes a multiplicative change by $\omega^k$ of the corresponding eigenvalue.
The behavior depends on the site's color.}
\label{fig:action_compound_operators}
\end{figure}

Here, we shall only consider perturbations which act locally on a site $\bi{i}$. 
A basis of the space of local unitary operators can be conveniently written down in terms of
the operators $X_{\bi{i}}$ and $Z_{\bi{i}}$ and their powers. Hermitian
combinations of these operators lead to the following general form of local uniform
perturbations
\begin{equation}
    \label{eq:HLM}
	H_{l,m} = -\sum_{\bi{i}} \left(h_{l,m} X_{\bi{i}}^{l}Z_{\bi{i}}^{m} +\mathrm{h.c.} \right)\, ,
\end{equation}
where $h_{l,m} \in \mathbb{C}$, and $(l,m) \in \mathbb{Z}_N^2$ with
$0\leqslant l\leqslant N/2$.
The action of the operator $X_{\bi{i}}^{l}Z_{\bi{i}}^{m}$ is displayed in
\fref{fig:action_compound_operators}. As can be inferred from this figure, the
perturbation violates the local conservation of charges (fluxes) when $m\neq 0$
($l\neq 0$) since it induces creation and annihilation of pairs of excitations
as well as hopping processes. However, it violates neither the conservation of
the total charge $\sum_s q_s$ modulo $N$, nor the conservation of the total flux $\sum_p q_p$
modulo $N$ [we recall that a star $s$ (a plaquette $p$) is said to
carry a charge $q_s$ (a flux $q_p$) if $A_s$ ($B_p$) has eigenvalue
$\omega^{q_s}$ ($\omega^{q_p}$), with $q_s$ and $q_p$ being defined modulo $N$.]

%
%
%
%

In fact, these conservation rules can be more conveniently rewritten as
conservation rules of ``unphysical'' total charge and flux, belonging to
$\mathbb{Z}$ and not to $\mathbb{Z}_N$ (they will also prove to be useful later on,
see \sref{subsec:self}).
Let us denote the ``unphysical'' charge (flux) at star $s$ (plaquette $p$) as
$\widehat{q}_s$ ($\widehat{q}_p$). They are related to the true/physical
charge (flux) via the following relations $q_s=\widehat{q}_s \mbox{ mod } N$
($q_p=\widehat{q}_p \mbox{ mod } N$).
They can be defined non ambiguously by ``fixing a gauge'' as we explain now
(having in mind a purely computational perspective). Given one eigenstate of
$H_\mathrm{TC}$, it is possible to choose any value for the ``unphysical''
charges and fluxes
(provided they yield the correct physical charges and fluxes). Then, the
full Hamiltonian $H_\mathrm{TC}+\sum_{l,m}H_{l,m}$ can be studied in the
subspace of eigenstates of $H_\mathrm{TC}$, spanned by the repeated action of
the perturbation. Each of these states can be assigned unique values of
$\widehat{q}_s$ and $\widehat{q}_p$ using rules shown in
\fref{fig:action_compound_operators}. It should be clear that
$\sum_s \widehat{q}_s$ and $\sum_p \widehat{q}_p$ are the same for all states in
the generated subspace, which means the Hamiltonian is block diagonal and the
total charge $\sum_s \widehat{q}_s$ and total flux $\sum_p \widehat{q}_p$ are
conserved integers.

%
\subsection{Simplest cases}\label{ssec:single_flavor}
%
Let us first discuss the case $l=0$ for which the perturbation acts only non trivially on charges (which are no longer locally conserved) whereas fluxes remain static gapped excitations that might be seen as 
sources of Aharonov-Bohm-like phases for the moving charges. Of course, the respective roles of fluxes and charges are exchanged if $l\neq 0$ and $m=0$.

In this work, we are interested in transitions between the topological
phase (existing for small enough perturbations) and the polarized phase expected for
$J=0$. To get a first idea about the associated physics, let us consider the
simplest perturbation corresponding to $(l=0,m=1)$, for which the
Hamiltonian of the perturbed toric code reads
\begin{equation}\label{eq:TC_SF}
\fl H_{\rm{TC}}+H_{0,1}= -J \sum_{s} \left( A_s^{\phantom{\dagger}} + A_s^{\dagger} \right) -J \sum_{p} \left( B_p^{\phantom{\dagger}} + B_p^{\dagger} \right) -h_Z \sum_{\bi{i}} \left( Z_{\bi{i}}^{\phantom{\dagger}} + Z_{\bi{i}}^{\dagger} \right) \, ,
\end{equation}
where we set $h_{0,1}=h_Z \in \mathbb{R}$ . Since we are interested in the low-energy properties and since 
$[H_{0,1},B_p]=0$ for all $p$'s, we only consider the
flux-free subspace where all $B_p$'s have eigenvalue $1$, in which the energy
of the fluxes is minimal.  Then, following the procedure discussed in
\cite{Trebst07,Hamma08} for the special case $N=2$, let us denote by $|q
\rangle_s$ the eigenstates of  $A_s$'s with eigenvalues $\omega^q$. From
\fref{fig:action_simple_operators} (right), one can check that the action of $Z_{\bi{i}}^{\phantom{\dagger}} + Z_{\bi{i}}^{\dagger}$ on a site ${\bi{i}}$ located between two neighboring stars $s$ and $s'$ is equivalent to $X_{s}^{\phantom{\dagger}} X_{s^{\prime}}^{\dagger} + X_{s}^{\dagger} X_{s^{\prime}}^{\phantom{\dagger}}$ where, following (\ref{eq:Definition_ZX_Operators}), we introduce the operator $X_{s}$ defined by 
$X_{s} |q \rangle_s=|q-1 \rangle_s$. Thus, defining $Z_{s}=A_s$ (such that $Z_{s}|q \rangle_s=\omega^q |q \rangle_s$), one can map Hamiltonian~(\ref{eq:TC_SF}) onto the $N$-state clock model in a transverse field \cite{Cobanera11}
\begin{equation}\label{eq:quClock}
	H_{\mathrm{clock}}=-2JN_p -J\sum_{s} \left( Z_s^{\phantom{\dagger}} + Z_s^{\dagger} \right) -h_Z \sum_{\langle s,s^{\prime}\rangle} 	\left( X_{s}^{\phantom{\dagger}} X_{s^{\prime}}^{\dagger} + X_{s}^{\dagger}X_{s^{\prime}}^{\phantom{\dagger}} \right) \, ,
\end{equation}
where $\langle s,s^{\prime}\rangle$ denotes nearest-neighbor stars $s$ and
$s^{\prime}$. The term $-2J N_p$ arises from the replacement of all
$B_p$-operators by their eigenvalue $1$ ($N_p$ denotes the total number of
plaquettes). It is important to stress that this mapping  preserves neither the
degeneracies of the energy levels (hence the topological order) nor the quantum
statistics. However, the zero-temperature phase diagrams of $H_{\rm{TC}}+H_{0,1}$ and $H_{\mathrm{clock}}$ are exactly the same.

Let us remark that the coupling term $h_Z$ in (\ref{eq:quClock}) stems from the local perturbation in (\ref{eq:TC_SF}) that can be either positive or negative.  When $h_Z>0$, the coupling between stars in  
$H_{\mathrm{clock}}$ is ferromagnetic whereas $h_Z<0$ leads to antiferromagnetic interactions. 
This distinction is irrelevant for even $N$ since, in this case (and for a bipartite lattice), one can always perform local unitary transformations which map $H_{\mathrm{clock}}$ onto a ferromagnetic model. By contrast, for odd $N$, one must distinguish between both signs that may lead to various types of transitions.

Unfortunately,  few results are available in the literature concerning the two-dimensional quantum clock model in a transverse field except for  $N=2$ (Ising model) where a second-order transition occurs, for $N=3$ (Potts model) where a weakly first-order is expected for $h_Z>0$ (see for instance \cite{Hamer92}), and for $N=4$ where the model is equivalent to two decoupled transverse-field Ising model \cite{Burnell11}.
In such a context, the second-order transition found in  \sref{sec:Z3F} for $N=3$ and $h_Z<0$  opens some interesting perspectives.

%
\subsection{Self-duality}
\label{subsec:self}
\begin{figure}[ht]
\centering
  \includegraphics[width=10cm]{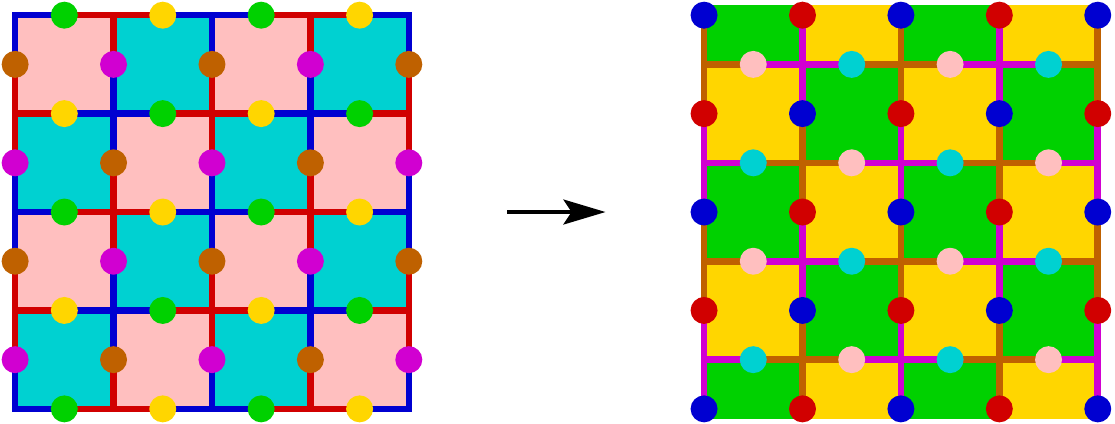} 
\caption{Transformation of the original lattice (left) to the dual lattice
(right). Plaquettes and stars become sites, and vice versa.}
\label{fig:self_duality_lattice}
\end{figure}

%
The $\mathbb{Z}_2$ toric code in a transverse field is known to be
self-dual \cite{Chen07,Vidal09_2}. We shall now show that this property still holds for
a general value of $N$, provided one chooses particular values of $l$ and $m$.
Let us consider the Hamiltonian $H_\mathrm{TC}+H_{l,m}$ [see 
(\ref{eq:TC_Hamilton}) and (\ref{eq:HLM})].
This Hamiltonian will be self-dual if its spectrum is symmetric (up to
degeneracies) under the exchange $J\leftrightarrow |h_{l,m}|$. 
Roughly speaking, this will be ensured provided  the ``roles'' of $A_s$ and $B_p$ operators can be
played by the operators $X_\bi{i}^l Z_\bi{i}^m$, once stars and plaquettes are
exchanged with sites, \textit{i.e.}, when considering the dual lattice
illustrated in \fref{fig:self_duality_lattice}. 

In the limiting cases, $J=0$ and $h_{l,m}=0$, self-duality imposes that  spectra of $H_{\rm TC}$ and  $H_{l,m}$ are the same (up to degeneracies). Setting $h_{l,m}=|h_{l,m}| {\rm e}^{{\rm i}\phi_{l,m}}$, this means that
${\rm e}^{{\rm i}\phi_{l,m}} X_\bi{i}^lZ_\bi{i}^m$ must have the same spectrum as  $A_s$ (or $B_p$). Since this spectrum is  $\{\omega^q,q\in\mathbb{Z}_N\}$, this leads to two constraints~: (i) $\left({\rm e}^{{\rm i}\phi_{l,m}} X_\bi{i}^l Z_\bi{i}^m\right)^N=\mathds{1}$ and (ii) $\left({\rm e}^{{\rm i}\phi_{l,m}} X_\bi{i}^lZ_\bi{i}^m\right)^n \neq\mathds{1}$, $\forall$ $n\in\{1,\ldots,N-1\}$.
Noting that $(X_\bi{i}^l Z_\bi{i}^m)^n=\omega^{l m \frac{n (n+1)}{2}} X_\bi{i}^{l n} Z_\bi{i}^{m n}$, the first constraint reads 
%
\begin{equation}
{\rm e}^{{\rm i}N\phi_{l,m}} \omega^{l m \frac{N (N+1)}{2}}=1\, ,
\label{eq:cond1}
\end{equation}
%
whereas the second one can be rephrased as
%
\begin{equation}
    l \,n \neq 0 \mbox{ mod } N \mbox { or } m\, n \neq0 \mbox{ mod } N \, ,\quad \forall n\in\{1,\ldots,N-1\}\, .
    \label{eq:cond2}
\end{equation}
%

\begin{figure}[ht]
\centering
  \includegraphics[width=6cm]{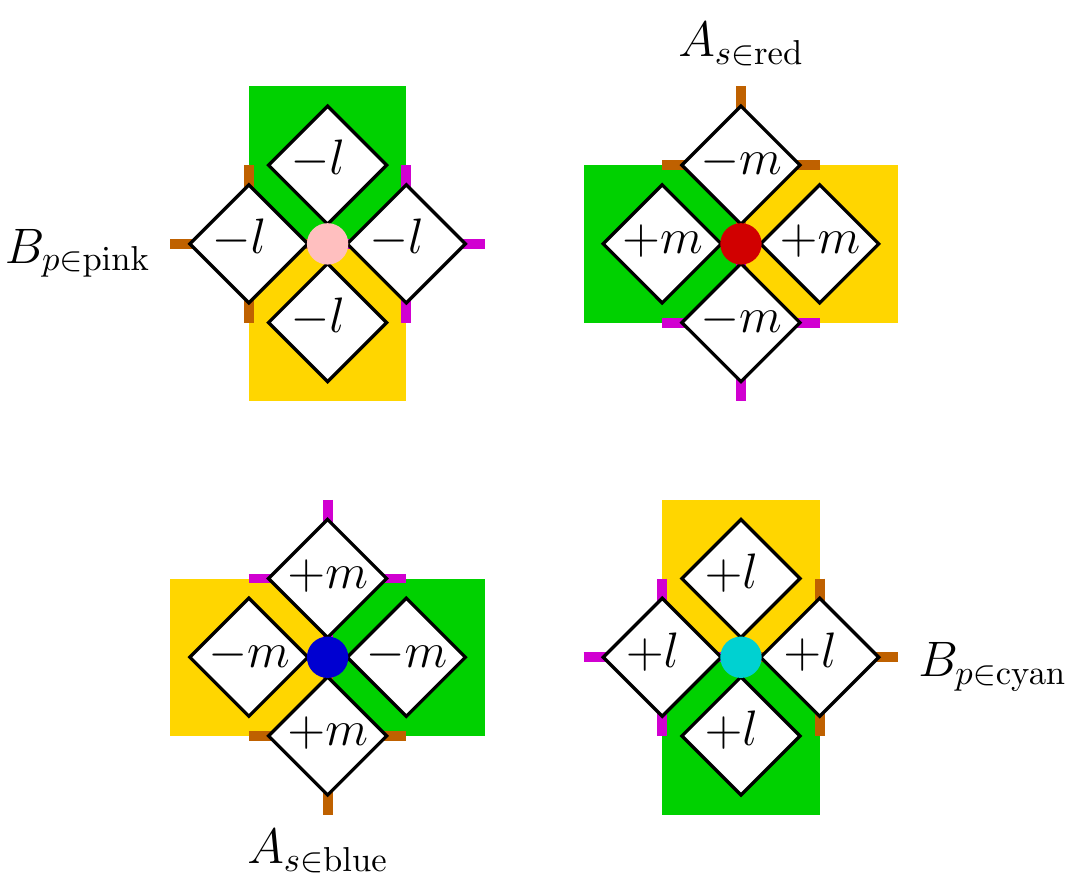} 
\caption{Illustration (on the dual lattice defined in the right part of 
figure~\ref{fig:self_duality_lattice} ) of the action of the operators $A_s$ and
$B_p$ on an eigenstate of $X_\bi{i}^l Z_\bi{i}^m$ operators for
all sites $\bi{i}$.
As in figure~\ref{fig:action_compound_operators}, a diamond on a star or a
plaquette, with a value $k=\pm l,\pm m$ denotes a multiplicative change by
$\omega^k$ of the corresponding eigenvalue. The behavior depends on the site's
color (so on the star and plaquette types on the original lattice).}
\label{fig:action_dual_operators}
\end{figure}

The third and most stringent condition is that the action of $H_{l,m}$ on
eigenstates of $H_\mathrm{TC}$, \textit{i.e.}, of all $A_s$ and $B_p$ operators
(see \fref{fig:action_compound_operators}), is the same as the action of
$H_\mathrm{TC}$ on eigenstates of $H_{l,m}$, \textit{i.e.}, of all
$X_\bi{i}^lZ_\bi{i}^m$ operators (see \fref{fig:action_dual_operators}).
After noticing that the numbers of minus signs, as well as their exact
positions, can be made the same in both figures by exchanging the roles of
$X_\bi{i}^lZ_\bi{i}^m$ and its hermitian conjugate on magenta and green sites
(or on yellow and brown sites), this last condition reads
%
\begin{equation}
    m=l \mbox{ or } m=N-l \, .
  \label{eq:cond3}
\end{equation}
%

For $N=2$, one recovers the model studied in \cite{Vidal09_2} (up to a factor of 2). Indeed, the only solution to the above equations is $l=m=1$ and $\phi_{1,1}=\pm \pi/2$. Thus, the perturbation reads $\pm 2 h_y \sum_\bi{i}\sigma_i^y$, while the toric code Hamiltonian can be simplified to $-2J(\sum_s A_s+\sum_p B_p)$ since the star and plaquette operators are Hermitian for $N=2$. Note that the present analysis also shows that the sign of the field term is irrelevant.

For the case $N=3$, to which the whole next section is devoted, condition (\ref{eq:cond1}) is fulfilled for ${\rm e}^{{\rm i}\phi_{l,m}}=1,\omega$ and 
$\omega^2$, whereas  (\ref{eq:cond2}) and (\ref{eq:cond3}) are fulfilled for 
$(l=1,m=1)$ and $(l=1,m=2)$. The six corresponding perturbations yield the same spectrum so that, for simplicity, we shall only consider 
$H_{1,1}$ with $\phi_{1,1}=0$. 

To conclude this discussion, let us show that self-duality in these models is also responsible for
additional symmetries, as was already observed for the special case
$N=2$ \cite{Vidal09_2}. Consider \fref{fig:action_compound_operators} with
$m=l$ (the same discussion holds for $m=N-l$). It is clear that the sum of the
numbers appearing in the diamonds, on diagonals or anti-diagonals, is either
$0$, $2l$ or $-2l$, which is always even. This allows one to define conserved
parity operators such as for example
$(-1)^{\sum_{s\in\mathrm{red}} \widehat{q}_s
+ \sum_{p\in\mathrm{pink}} \widehat{q}_p}$,
where the sum runs overs red stars and pink plaquettes forming a given
diagonal. In order for this parity operator to be conserved when $N$ is odd, one
has to consider the charge and flux numbers $\widehat{q}_s$ and $\widehat{q}_p$ belonging to $\mathbb{Z}$, introduced at the end of \sref{sec:sub:gen_struct_pert} instead of $q_s$ and $q_p$ that belong to $\mathbb{Z}_N$.
Of course, a dual discussion can be given by working on the dual lattice and in
the eigenbasis of $X_\bi{i}^lZ_\bi{i}^m$ operators. As in the $\mathbb{Z}_2$
toric code in a transverse field, the conservation of these parity operators
constrains the dynamics, and the model displays dimensional reduction. This dimensional reduction was originally discussed in the Xu-Moore model \cite{Xu04,Xu05,Nussinov05} which has the same spectrum as the $\mathbb{Z}_2$ toric code in a transverse field. A detailed discussion about these issues can be found in \cite{Nussinov11}.

%
%
%
%

%
%
%
%

\section{Perturbing the $\mathbb{Z}_3$ toric code}
\label{sec:Z3F}
\subsection{Model}

In this section, we focus on the case $N=3$ and we study the robustness of  the $\mathbb{Z}_3$ toric code with respect to simple perturbations. To this aim, and for the sake of simplicity, we consider the following Hamiltonian
\begin{eqnarray}\label{Eq:Hamiltonian_Z3TCF}
\fl	 H(h_X,h_\perp,h_Z) &=&H_{\rm TC}+H_{1,0}+H_{1,1}+H_{0,1},\\
	 &=&-\frac{1}{3} \sum_{s} \left(A^{\phantom{\dagger}}_s+A^{\dagger}_s\right)-\frac{1}{3} \sum_{p} \left(B^{\phantom{\dagger}}_p+B^{\dagger}_p\right)-h_X\sum_{\bi{i}} \left(X^{\phantom{\dagger}}_{\bi{i}}+X^{\dagger}_{\bi{i}}\right) \nonumber \\
	\fl &&-h_{\perp}\sum_{\bi{i}} \left(X_{\bi{i}}Z_{\bi{i}}+Z_{\bi{i}}^{\dagger}X_{\bi{i}}^{\dagger}\right) -h_Z\sum_{\bi{i}} \left(Z^{\phantom{\dagger}}_{\bi{i}}+Z^{\dagger}_{\bi{i}}\right)  , 
\end{eqnarray}
where we choose $J=1/3$ in order to set the elementary excitation gap of the
unperturbed Hamiltonian $H_{\rm TC}$ to unity. In addition, we restrict our discussion to real parameters and set 
$h_{1,0}=h_X$, $h_{1,1}=h_\perp$,  $h_{0,1}=h_Z$. 
For $N=2$, this Hamiltonian (with the proper phase factor for $H_{1,1}$ discussed in \sref{subsec:self}) corresponds to the $\mathbb{Z}_2$ toric code in a uniform magnetic field studied in \cite{Dusuel11} so that this choice is well suited for a comparison between both systems.
In the following, we adopt a language similar to that used for $N=2$. Thus,
$H_{1,0}$ and $H_{0,1}$ ($H_{1,1}$) will be considered as ``parallel" (``transverse") perturbations. 

The determination of the full three-dimensional phase diagram of Hamiltonian
$H(h_X,h_\perp,h_Z)$ is a difficult   problem. Therefore, as was done initially
for the $\mathbb{Z}_2$ case \cite{Vidal09_1,Vidal09_2}, we will study parallel
and transverse cases separately, but let us first discuss the methods used.

%
\subsection{Methods}
\label{ssec:methods}
As already discussed, the system has to undergo phase transitions when perturbed with local operators. As in the perturbed $\mathbb{Z}_2$ toric code, one expects first- and second-order transitions in the phase diagram \cite{Dusuel11}. In order to analyze the breakdown of the topological phase, we  combine the pCUT method and the variational iPEPS algorithm. This approach is motivated by the fact that the pCUT gives reliable estimates for second-order phase transitions  whereas the iPEPS algorithm, as a variational tool, is especially sensitive to first-order transitions.

%
\subsubsection{pCUT}
%
The method of continuous unitary transformations has been introduced in reference \cite{Wegner94} and general aspects of its perturbative variant pCUT can be found in reference \cite{Knetter00}. In what follows, we focus on points that are specific to the application of the pCUT method to a topologically-ordered phase \cite{Vidal09_1,Vidal09_2,Dusuel11}.

To apply the pCUT, it is essential that the unperturbed Hamiltonian (here
$H_{\rm TC}$) possesses an equidistant spectrum \footnote{We would like to stress that
the spectrum of the $\mathbb{Z}_N$ model studied in this paper is only
equidistant for $N=2,3,4$, so that one cannot use this machinery for
other values of $N$. However, it could be applied to Kitaev's $\mathbb{Z}_N$ toric code \cite{Kitaev03}, whose
spectrum is equidistant for all $N$.} that is bounded from below \cite{Knetter00}. These two constraints are satisfied in the $\mathbb{Z}_3$ toric code as long as  gaps of charges and fluxes are identical. In this case, one can interpret the toric code as a counting operator $Q$ of charges and fluxes in the system
\begin{eqnarray}\label{Eq:pCUT_Q}
	H_{\rm TC} &=&-\frac{2}{3} \left( N_s + N_p \right) + Q \, ,
\end{eqnarray}
where $N_s$ ($N_p$) denotes the total number of stars (plaquettes). Therefore
the constant term represents the ground-state energy (remember that we set
$J=1/3$). 

It is then possible to rewrite the local perturbations as $\sum_n T_n$, where
$T_n$ changes the particle number in the system by $n$, \textit{i.e.},
$[Q,T_n]=n\, T_n$.
The pCUT maps, order by order in the perturbation, the Hamiltonian $H$
onto an effective Hamiltonian $H^{\mathrm{eff}}$, unitarily equivalent to $H$ (same spectrum but different eigenstates), that reads as
follows in the eigenbasis of the \textit{bare} Hamiltonian $H_\mathrm{TC}$~:
\begin{eqnarray}\label{eq:eff_Ham}
 \fl H^{\mathrm{eff}} = -\frac{2}{3} \left( N_s + N_p \right) + Q\ +\
 \sum_{k=1}^{\infty} \quad \sum_{m_1+\ldots+m_k=0} C\left(m_1,\ldots,m_k\right)
 T_{m_1} \ldots T_{m_k} \, .
\end{eqnarray}
As explained in \cite{Knetter00}, the coefficients $C(m_1,\ldots,m_k )$ are
model-independent rational numbers.
 An essential property of
the effective Hamiltonian is that $\left[ H^{\mathrm{eff}},Q\right]=0$. 
As a consequence, the number of quasiparticles (QPs) in the system, {\it i.e.}, eigenstates of $Q$, is a good quantum number.
In the perturbed toric code, QPs are dressed anyons adiabatically connected to the corresponding bare charges and fluxes.

To determine the zero-temperature phase diagram, we focus on the low-energy spectrum of
$H^{\mathrm{eff}}$. Essentially, one must study 
$H^{\mathrm{eff}}$ in the 0-QP subspace (to compute the ground-state energy) and in the 1-QP subspace (to obtain the low-energy gap~$\Delta$). We emphasize that computing the low-energy gap from the 1-QP subspace is meaningful as long as there are no bound states with lower energies. This is a working hypothesis that is crucial in what follows.
As discussed in \sref{sec:gs}, for open boundary conditions, the ground-state is
non degenerate and the ground-state energy is nothing but the expectation value
of $H^{\mathrm{eff}}$ on the ground state of the bare Hamiltonian
%
%
\begin{equation}
\label{eq:eff_Ham_E0}
 	E_0=\left\langle \mathrm{gs} \right| H^{\mathrm{eff}} \left| \mathrm{gs} \right\rangle \, .
\end{equation} 
%
%

The structure of the 1-QP subspace is less trivial. Indeed, for $N=3$ there
are four different kinds of excitations~: charges ($q_s=1$) and anti-charges
\mbox{($q_s=-1=2 \mbox{ mod } 3$)} living on stars, as well as fluxes ($q_p=1$) and anti-fluxes
\mbox{($q_p=-1=2\mbox{ mod } 3$)} living on plaquettes.
The associated subspaces are \textit{not} connected by $H^\mathrm{eff}$
because of the symmetry ensuring that the total charge and total flux are
conserved (modulo $N=3$ when working with the physical charge and flux).
Thus $H^\mathrm{eff}$ is a 1-QP hopping Hamiltonian in each of these sectors,
and is therefore easily diagonalized, once the hopping amplitudes are
determined. One then obtains dispersion relations for the four kinds of
excitations, which only give two different energies because charges and
anti-charges, as well as fluxes and anti-fluxes, play symmetric roles. Finally,
one can compute the gap $\Delta$ as the minimum over all momenta of these two
energy bands.

Let us point out that the major challenge, in both the 0-QP and 1-QP
sectors, lies in the computation of matrix elements of $H^\mathrm{eff}$. Indeed, one has to take into
account the non-trivial braiding phases coming from virtual fluctuations of the excitations.
Although the method yields results in the thermodynamical limit, the linked-cluster theorem (see for example
\cite{Knetter03,Dusuel10} and references therein) allows one to compute the hopping amplitudes by considering finite-size clusters (albeit one needs a growing number of them when the order of the perturbation increases).

At the end of the day, one obtains a high-order series expansion of the
ground-state energy $E_0$ and of the 1-QP gap $\Delta$. The extrapolation of $\Delta$
with standard resummation techniques (see {\it e.g.} \cite{Oitmaa06}) allows a reliable determination of second-order phase transition points and thus of the boundaries of the topological phase
(assuming that bound states of elementary QPs are not relevant and do not have
an energy smaller than that of a single QP).

Unfortunately, as already stated, series expansions are not adapted to detect possible
first-order phase transitions (in particular when having series in one phase
only) so that one needs a complementary tool which we now describe.
%
\subsubsection{iPEPS}
\label{subsubsec:iPEPS}
%
The so-called iPEPS algorithm \cite{Jordan08} is a variational method which, as such,
is aimed at approximating the ground state of two-dimensional quantum lattice
systems by employing a tensor-network approach.
Details about this method have already been extensively discussed in the literature \cite{Corboz10, Jordan08, Jordan11, Orus09_2, Corboz11}. For completeness, though, we explain some of the basic features of the algorithm, focusing on  those that are relevant for the study of the perturbed $\mathbb{Z}_3$ toric code. 

\begin{figure}
\centering
  \includegraphics[width=6cm]{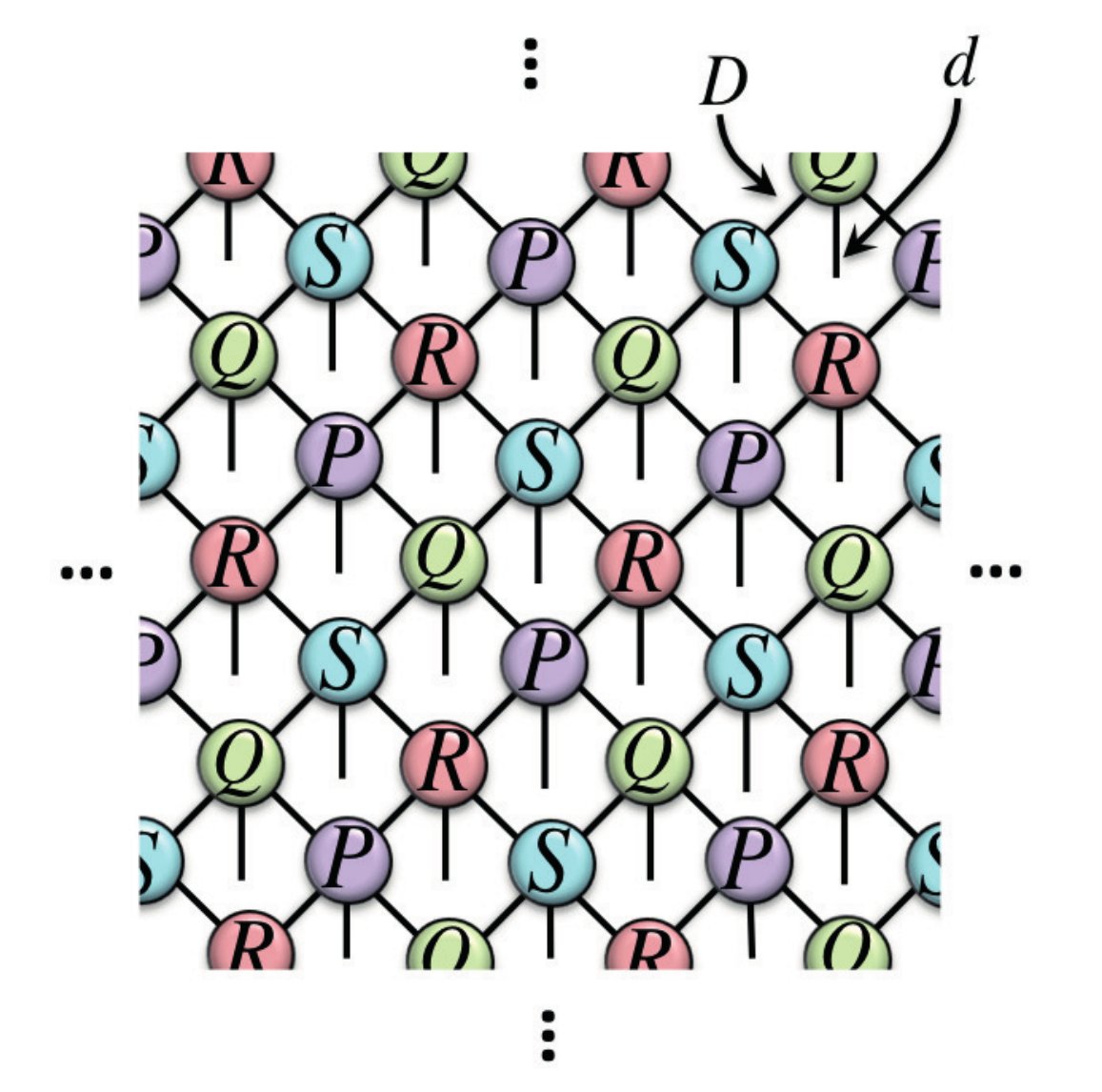} 
\caption{Schematic representation of the (translation-invariant) PEPS considered in this study with four different tensors ($P$, $Q$, $R$, and $S$) per unit cell. Tensors are represented by circles, and their indices by lines. Lines that connect different circles correspond to bond indices shared by two tensors and can take up to $D$ different values. Open lines correspond to physical indices, which take $d=N=3$ values (dimension of the local Hilbert space on each site).}
  \label{fig:PEPS}
\end{figure}

In the iPEPS algorithm, the quantum state $|\Psi\rangle$ of the infinite square lattice is
represented by a projected entangled pair state (PEPS)
\cite{Jordan08,Verstraete04}. In the present problem, we choose a  PEPS
with four tensors denoted $P$, $Q$, $R$,  and $S$  per unit cell (see
\fref{fig:PEPS}).  Each of these tensors depends on $\Or(d\, D^4)$ complex
coefficients, where $d=N=3$ is the dimension of the local Hilbert space at each site, and $D$ is the so-called bond dimension of the PEPS. This bond dimension controls the maximum amount of entanglement carried by the PEPS wave function and, consequently, the accuracy of the ansatz. Following the discussion in \cite{Verstraete06} for $N=2$, one can show that the ground state of the (non-perturbed) $\mathbb{Z}_N$ toric code is a $D=N$-PEPS. Obviously, for $J=0$ the fully-polarized ground state is also a (trivial) PEPS ($D=1$). Consequently, at least in these two limiting cases, PEPS are exact ground states of the Hamiltonian.

In practice, for a given $D$, the goal is to find the coefficients of tensors $P$, $Q$, $R$ and $S$ that best approximate the ground state of $H$. These coefficients can be determined by an imaginary-time evolution driven by the Hamiltonian, since
\begin{equation}
	|\Psi_{\mathrm{gs}}\rangle = \lim_{\tau \rightarrow \infty} \frac{e^{-\tau H }|\Psi_0\rangle}{||e^{-\tau H} |\Psi_0\rangle||}\, ,
\end{equation}
where $|\Psi_{\mathrm{gs}}\rangle$ is the ground state of $H$ and $|\Psi_0\rangle$ is any initial state that has a non-vanishing overlap with the ground state. The approximation of this evolution is performed in a similar way as explained, for instance, in \cite{Jordan08,Orus09_2}.
\begin{enumerate}
\item\label{it:iPEPS1} The whole evolution is splitted into small imaginary-time steps $\delta \tau$ by using a Suzuki-Trotter expansion of the evolution operator $e^{-\tau H}$. More precisely, writing the Hamiltonian as a sum of four-body terms $h^{[ \bi{i} \bi{j}  \bi{k} \bi{l} ]}$,
\begin{equation}
    H = \sum_{\bi{i} \bi{j}  \bi{k} \bi{l}} h^{[ \bi{i} \bi{j}  \bi{k} \bi{l} ]}\, ,
\end{equation}
we consider, at each time step, the action of the four-site operator 
\begin{equation}
	g^{[ \bi{i} \bi{j}  \bi{k} \bi{l} ]} \equiv e^{-\delta \tau h^{[ \bi{i} \bi{j}  \bi{k} \bi{l} ]}}  .
\end{equation}
\item\label{it:iPEPS2} At each imaginary-time step the state is approximated by some PEPS with the considered structure and bond dimension $D$. For instance, if at step $\tau$ we have a PEPS $|\Psi(\tau) \rangle$, then the evolved state $|\widetilde{\Psi}(\tau + \delta \tau) \rangle \equiv g^{[ \bi{i} \bi{j}  \bi{k} \bi{l} ]} |\Psi(\tau)\rangle$ is also  approximated by a new PEPS $|\Psi(\tau + \delta \tau) \rangle$ with the same structure. Practically, this approximation is achieved by minimizing the distance $|| |\widetilde{\Psi}(\tau + \delta \tau) \rangle - |\Psi(\tau + \delta \tau) \rangle||^2$ with respect to the coefficients of tensors $P$, $Q$, $R$ and $S$ of the new PEPS. In our case, we have carried out this minimization simultaneously over the four tensors by using a standard conjugate-gradient algorithm.
\end{enumerate}

Quite importantly, step (ii) as well as the evaluation of expectation values of local observables involves the contraction of an infinite two-dimensional tensor network. This contraction can be approximated by various schemes \cite{Jordan08,Orus09_2,Gu08}. Here, we choose the Directional Corner Transfer Matrix Approach introduced in \cite{Orus09_2} that can be easily adapted to deal with different types of two-dimensional tensor networks \cite{Corboz11}, including those considered here. An important parameter in these manipulations is the so-called bond dimension of the environment $\chi$ which controls the accuracy of the approximations involved at this step.

Thus, according to the discussion above, there are four possible sources of error in the iPEPS algorithm.

\begin{enumerate} 
\item{The size and the shape of the considered unit cell. The error is reduced when the unit cell gets larger.}
\item{The bond dimension $D$ of the PEPS. The error is reduced when $D$ gets larger.}
\item{The imaginary-time step $\delta \tau$. The error is reduced when $\delta \tau$ gets smaller.}
\item{The bond dimension of the environment $\chi$. The error is reduced when $\chi$ gets larger.}
\end{enumerate}

In this study, we have fixed $D=3, \chi=30$, $\delta \tau = 10^{-3}$, and the aforementioned ``four-site" unit cell. We checked that the increase of precision obtained by varying the values of the parameters is within the error bars obtained by the pCUT approach. Therefore, this choice of parameters turns out to be sufficient for our purposes.

%
\subsubsection{pCUT+iPEPS}

To determine the boundaries of the topological phase, we combine results from pCUT and iPEPS algorithm as explained below. To simplify the discussion, let us assume that we have only one control parameter $h>0$.

Let us recall that a second-order transition is associated to the closure of the low-energy gap. Assuming that the relevant gap comes from the 1-QP sector, the critical point $h^{\rm c}$ is then defined by $\Delta\left(h^{\rm c}\right)=0$. As already mentioned, this point can be efficiently computed with the pCUT method by extrapolating the high-order series expansion of $\Delta$. Typically, with the maximum orders reached in this study, one can determine $h^{\rm c}$ with a relative precision of the order $10^{-2}$-$10^{-3}$. However, if some level crossing occurs, one faces a first-order transition that cannot be captured by the criterion $\Delta=0$.
This is why it is crucial to use the iPEPS algorithm to compute, variationally, the ground-state energy. 
Indeed,  denoting by $e_0^{\rm pCUT}$ and $e_0^{\rm iPEPS}$ the ground-state
energies calculated by both methods and assuming the existence of a point $h^{\star}$ where $e_0^{\rm iPEPS}<e_0^{\rm pCUT}$,  two cases must be considered. If $h^{\star}>h^{\rm c}$, it means that the iPEPS algorithm does not detect any level crossing for the ground state before the critical point, and hence a second order transition is likely taking place at $h^{\rm c}$.
By contrast, if $h^{\star}<h^{\rm c}$, then it means that a level crossing definitely occurs before the critical point $h^{\rm c}$ and we conclude that a first-order transition takes place at $h^{\star}$. 

Of course, this reasoning would be exact if (i) one would have an infinite
precision in both methods which is clearly not the case and (ii) no bound-state with lower energy exists (see discussion above).
The accuracy of this
combined pCUT+iPEPS approach is limited by several sources. First, the series
expansion is performed up to a finite maximum order and the error of resummation
schemes like the dlog-Pad\'{e} extrapolation is hard to quantify. Second, the
variational iPEPS algorithm is limited by the values of the parameters $D$, $\chi$, $\delta \tau$ as well as the structure of the tensor network itself (see \sref{subsubsec:iPEPS}). Additionally, it is numerically challenging to extract the global minimum of the variational ground-state energy. 
Reasonably, one can state that the combined pCUT+iPEPS approach works well as
long as the error bars of both methods when determining the values $h^{\star}$ and $h^{\rm c}$ are small compared to the difference $\left| h^{\star} - h^{\rm c} \right|$. As explained above, in the present work, this relative error on $h^{\star}$ and $h^{\rm c}$ is of the order $10^{-2}$-$10^{-3}$.

\subsection{Results}\label{ssec:Results}

Let us now present our results concerning the perturbed  $\mathbb{Z}_3$ toric code $H(h_X,h_\perp,h_Z)$ [see (\ref{Eq:Hamiltonian_Z3TCF})] for several limiting cases.

%
\subsubsection{The case $h_X=h_\perp=0$ }
\label{sssec:Potts}
%

As already mentioned in \sref{ssec:single_flavor}, the low-energy physics of
the perturbed toric code $H(0,0, h_Z)$ corresponds to the $N$-state clock model
in a transverse field (\ref{eq:quClock}). For $N=3$, this model is also
equivalent to the three-state Potts model in a transverse field. The
extension of the topological phase can therefore be obtained by directly
analyzing this model, which is simpler since one only has to consider charge degrees of freedom that live on stars of the square lattice. Indeed, remind that fluxes are frozen (conserved) and even absent in the low-energy sector.

As a first step, apart from the pCUT+iPEPS analysis, let us perform a
standard mean-field calculation that, as we shall see, already captures the main
qualitative aspects of the phase diagram although it relies on a trivial (non-entangled) variational state. To this aim, let us consider the following trial wave function 
\begin{eqnarray}
	\fl \left|\Psi\right\rangle = \bigotimes_{s\in \mathrm{blue}} \left| \psi_{\mathrm{b}}\right\rangle_{s^{\phantom{\prime}}} \bigotimes_{s\in \mathrm{red}} \left| \psi_{\mathrm{r}}\right\rangle_{s}, \ \mathrm{with}\
	\left| \psi_{\mathrm{b,r}}\right\rangle_{s^{\phantom{\prime}}}=a_{\mathrm{b,r}} \left| 0 \right\rangle_s + b_{\mathrm{b,r}} \left| 1 \right\rangle_s + c_{\mathrm{b,r}} \left| 2 \right\rangle_s \, ,
\end{eqnarray}
where the coefficients $a_{\mathrm{b,r}}$, $b_{\mathrm{b,r}}$, and
$c_{\mathrm{b,r}}$ are chosen such that the local wave functions $\left|
\psi_{\mathrm{b,r}}\right\rangle_{s^{\phantom{\prime}}}$ are normalized and
minimize  $\langle \Psi |H| \Psi \rangle$. The introduction of different
wave functions on red (r) and blue (b) stars is needed to accomodate with both
ferromagnetic order and anti-ferromagnetic order (expected for $h_Z>0$ and
$h_Z<0$ respectively). For the clock model, the sublattice magnetization
\begin{eqnarray}
 	M_{\mathrm{b},\mathrm{r}}\left(h_Z\right)=\left(\left\langle \psi_{\mathrm{b},\mathrm{r}} \right| X_s \left| \psi_{\mathrm{b},\mathrm{r}} \right\rangle\left\langle \psi_{\mathrm{b},\mathrm{r}} \right| X_s^{\dagger} \left| \psi_{\mathrm{b},\mathrm{r}} \right\rangle \right)^{\frac{1}{2}} \, ,
\end{eqnarray}
is a proper order parameter. The topological phase of the perturbed  $\mathbb{Z}_3$ toric code corresponds to the disordered (symmetric) phase of the three-state clock model characterized by $M_{\mathrm{b},\mathrm{r}}=0$. 
This phase is obtained for $\left|h_Z\right|\ll J=1/3$. By contrast, an ordered (broken) phase with $M_{\mathrm{b},\mathrm{r}}\neq 0$ is expected for large perturbations $\left|h_Z\right|\gg J=1/3$. Once again, let us emphasize that the sign of $h_Z$ is important for $N=3$. Indeed,  for large positive $h_Z$, a polarized phase (with all stars in the same state) is stabilized, whereas for large negative $h_Z$ a ``staggered" order arises.

\begin{figure}[ht]
\centering
  \includegraphics[width=12cm]{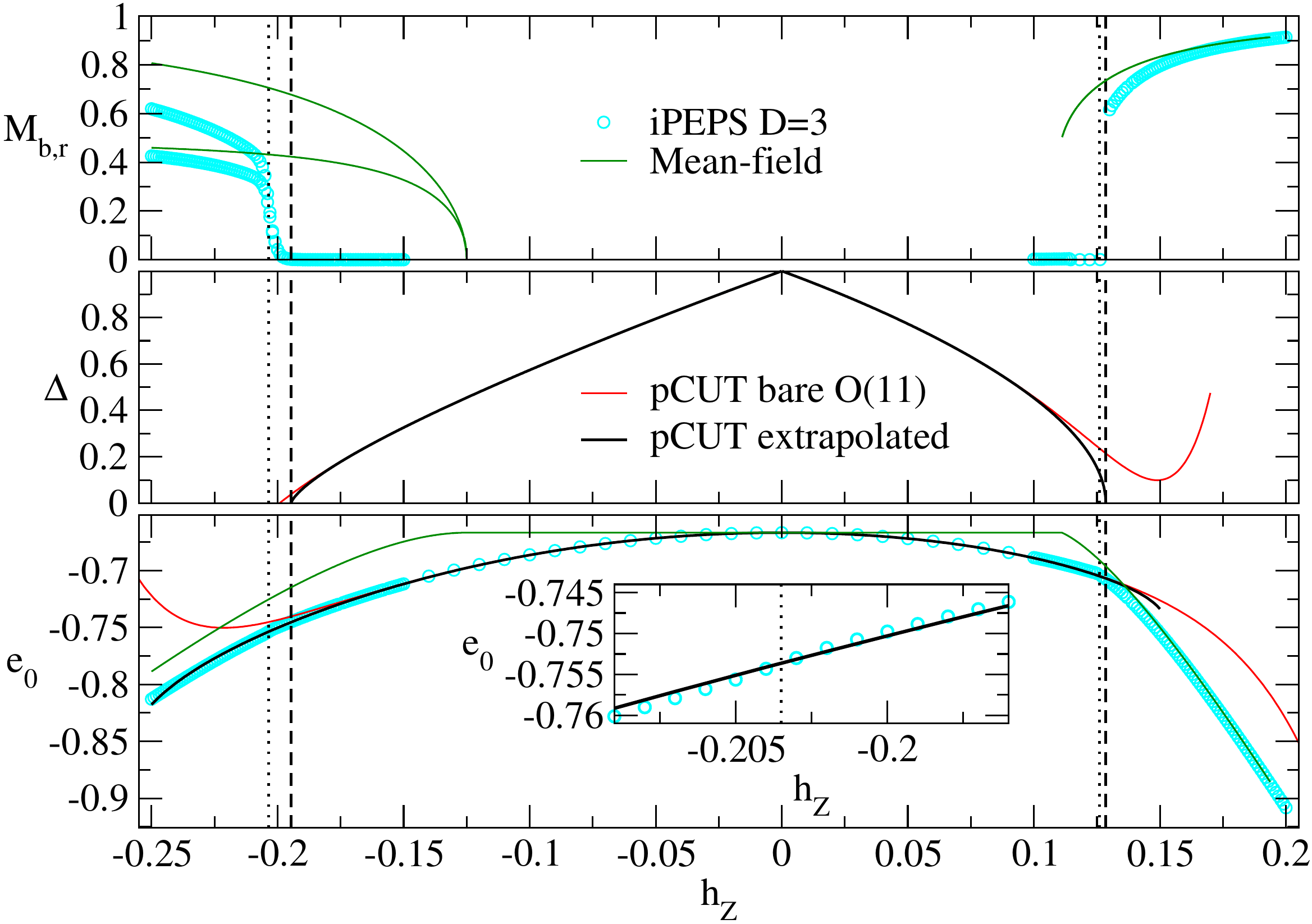} 
  \caption{Comparison of pCUT, iPEPS, and mean-field results for the $N=3$ clock (Potts) model  in a transverse field $h_Z$ ($J=1/3$). At  order 11, all dlog-Pad\'{e} approximants agree within the width of the black lines. Vertical dashed lines indicate the position of  $h^{\rm c}_Z$ defined by  $\Delta (h^{\rm c}_Z)=0$. Vertical dotted lines indicate the position of  $h^{\star}_Z$ beyond which $e_0^{\rm iPEPS}<e_0^{\rm pCUT}$. Upper panel~: sublattice magnetizations $M_{\mathrm{b,r}}$ as a function of $h_Z$. Central panel~: 1-QP gap $\Delta$ as a function of $h_Z$.  Lower panel~: ground-state energy per site $e_0$ as a function of $h_Z$.  Inset~: zoom of $e_0$ close to the critical point  for negative $h_Z$.}
  \label{fig:Potts}
\end{figure}

Our results for the mean-field order parameter are summarized in the upper panel of \fref{fig:Potts}. For $h_Z>0$, we find a first-order transition at $h_Z^{1,{\rm MF}}=1/9$. At this point, the order parameter jumps discontinuously from 0 to  1/2. For $h_Z>h_Z^{1,{\rm MF}}$, the system is uniformly polarized ($M_{\mathrm{r}}=M_{\mathrm{b}}$) and $\displaystyle{\lim_{h_Z \to +\infty} M_{\mathrm{b,r}}=1}$, as it should.
For  $h_Z<0$,  we obtain a continuous quantum phase transition at \mbox{$h_Z^{2,{\rm MF}}=-1/8$}. As discussed above,  the ordered phase found for $h_Z<h_Z^{2,{\rm MF}}<0$ spontaneously breaks the translation symmetry of the system, \textit{i.e.}, $M_{\mathrm{b}} \neq M_{\mathrm{r}}$. 
In the mean-field approximation,  both (sublattice) order parameters $M_{\mathrm{r}}$ and $M_{\mathrm{b}}$ vanish as $(h_Z^{2,{\rm MF}}-h_Z)^{\beta^{\rm MF}}$ with $\beta^{\rm MF}=1/2$. 
Furthermore, as can be shown by a simple first-order perturbation theory in $J/h_Z$,  one has $\displaystyle{\lim_{h_Z \to -\infty} M_{\mathrm{b}}=1}$ and$\displaystyle{\lim_{h_Z \to -\infty} M_{\mathrm{r}}=1/2}$ (or vice versa).

To go beyond this mean-field analysis and to obtain more quantitative results, let us now turn to the pCUT+iPEPS analysis. As the Hamiltonian (\ref{eq:quClock}) only contains  one-site and two-site terms, the evaluation of effective matrix elements in the pCUT calculation can be performed using a full graph decomposition \cite{Oitmaa06,Yang10} that allowed us to reach order 11 (see \ref{ssec:Details_Potts}). The iPEPS algorithm also considerably benefits  from the absence of four-site terms in $H_{\rm clock}$.
As can be seen in Fig.~\ref{fig:Potts}, the PCUT+iPEPS method is in qualitative agreement with the mean-field treatment since it also predicts a first-order transition for  $h_Z>0$ and a second-order transition for $h_Z<0$.
However, for $h_Z>0$ where our results match those given in \cite{Hamer90} for the Potts model (up to a rescaling), we find $h^{\star}_Z \simeq 0.126$ and $h^{\rm c}_Z\simeq 0.129$. Thus, within the combined pCUT+iPEPS scheme, we are led to conclude that a (weakly) first-order phase transition occurs at 
$h_Z^{\star} \simeq 0.126$  in agreement with the results given in \cite{Hamer92,Jordan11}. As can be seen in Fig.~\ref{fig:Potts} (upper panel), the calculation of the order parameter using the iPEPS algorithm confirms a first-order behavior (jump of the magnetization). Note also that the relative difference between $h^{\star}_Z$ and $h_Z^{1,{\rm MF}}$ is about  $ 12 \%$. 

The situation is different for $h_Z<0$. In this case, we find  $h^{\star} \simeq -0.204$ and $h^{\rm c}_Z \simeq -0.195$ (see \tref{tab:dlogpotts} for more details).  Let us stress that we observed a very good agreement between iPEPS and PCUT calculations with a relative error between both ground-state energies smaller than $10^{-4}$ for $h^{\rm c}_Z<h_Z<0$. We therefore conclude that, within our scheme,  a second-order phase transition occurs at $h^{\rm c}_Z \simeq -0.195$. In this parameter region, the discrepancy between $h^{\rm c}_Z$ and $h_Z^{2,{\rm MF}}=-1/8$ is larger than $35\%$.
To the best of our knowledge, this second-order phase transition has never
been discussed in the literature and it is therefore very desirable to further
characterize its universality class. Unfortunately, the iPEPS calculation of the
critical exponent $\beta$ associated to the order parameter is known to be
specially sensitive to finite-$D$ effects. As shown in the one-dimensional Ising
model in a transverse field \cite{Liu10}, this exponent also eventually reaches
its mean-field value 1/2 for any finite $D$ when approaching the critical point. Nevertheless, the exact value (1/8 in
the latter problem) can be observed in a field range near the critical point whose size increases with $D$. In a two-dimensional system, it is very  difficult to increase $D$ in order to perform a similar study. For the problem at hand, our fixed bond dimension $D=3$ only allowed us to see an exponent different from $\beta^{\rm MF}=1/2$ in a very small region, and this was not sufficient to determine a reliable value.

Alternatively, dlog-Pad\'e extrapolations of high-order series expansion of $\Delta$ allows one to determine the exponent driving the closure of the gap at the critical point. More precisely, denoting $z$ the dynamical exponent and $\nu$ the correlation length exponent, one has $\Delta\sim |h-h^{\rm c}|^{z\, \nu}$  for $h$ near $h^{\rm c}$ (see {\it e.g.} \cite{Dutta11}). At order 11, we found $z\, \nu \simeq 0.71$ but, as can be seen in \tref{tab:dlogpotts}, it is clear that this value is poorly converged. 
However, to roughly estimate the error, one may draw a parallel with the $N=2$
problem (\textit{i.e.}, the Ising model in a transverse field) for which series expansion of the gap have been computed up to order 13 in \cite{He90}. There, using the order 11 results, one gets an exponent $z \, \nu \simeq 0.645$ which only differs by a few percent from the commonly accepted values $z=1$ and $\nu=0.630(1)$. Hopefully, for $N=3$, we are also close from the asymptotic value but one clearly needs a more quantitative study to clarify the nature of this quantum phase transition. From that respect, Monte-Carlo simulations could provide valuable insights.

%
\subsubsection{The case $h_\perp=0$ }\label{sssec:Higgs}
%
Let us now turn to the case when both perturbations $H_{0,1}$ and $H_{1,0}$ are present so that neither charges nor fluxes are locally conserved anymore.  Thus, one has to treat charges and fluxes at the same level and to carefully take their mutual statistics into account in the virtual braiding processes. This constraint strongly reduces the maximum order that can be reached in the pCUT calculation. Instead of order 11, we only computed $e_0$ and $\Delta$ up to order 7 for this case (see \ref{ssec:Details_Potts}). The iPEPS calculation becomes more involved as well since one now has to deal  with 4-body interactions  (instead of 2-body interactions in the three-state clock model). In other words, our results are less accurate when a more complex perturbation is considered, as expected.

\begin{figure}[ht]
\centering
  \includegraphics[width=12cm]{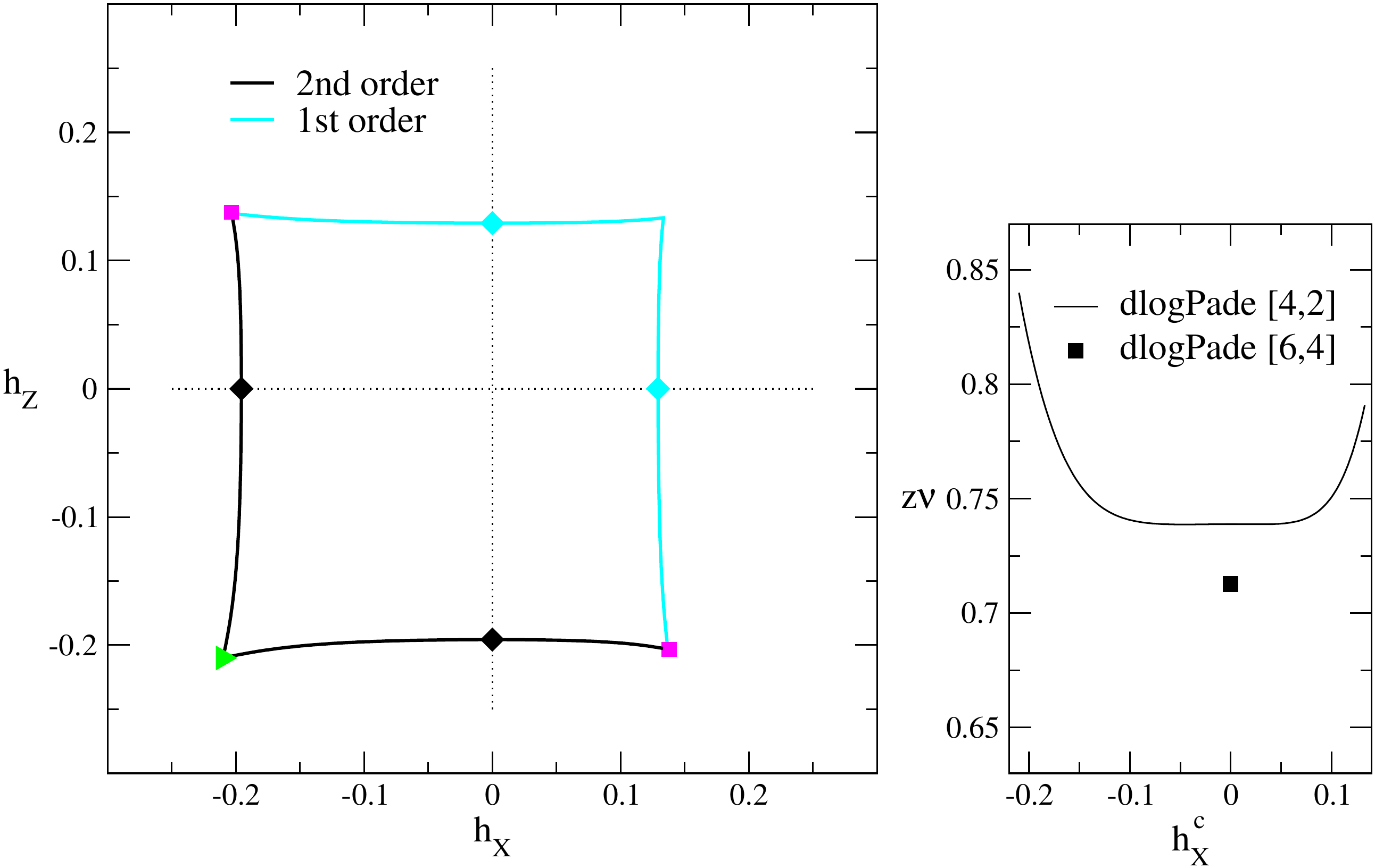} 
  \caption{Left~: zero-temperature
  phase diagram of $H(h_X,0,h_Z)$ obtained by the pCUT+iPEPS approach. Cyan
  (black) lines denote first-order (second-order) phase transitions. The
  first-order (second-order) phase transitions for the ferromagnetic
  (anti-ferromagnetic) three-state clock model in a transverse field are marked
  by cyan (black) diamonds. Magenta squares locate the crossings between first-
  and second-order transition lines while the crossing of the two second-order
  lines is taking place at the green triangle.
  Right~: plot of the gap exponent $z\, \nu$ as
  a function of $h^{\rm c}_X$ along the horizontal second-order phase transition
  line (see left part). The black solid line is the result obtained from the
  order seven expression of the 1-QP gap. The black square is our best estimate
  of $z\, \nu$ for the simpler case of one parallel perturbation using the order
  eleven series expansion obtained for the three-state Potts model in a
  transverse field (see \sref{sssec:Potts}).}
  \label{fig:Higgs}
\end{figure}

In \fref{fig:Higgs} (left), we display the phase diagram obtained by combining pCUT and iPEPS algorithm following the same procedure as previously but for arbitrary directions in the $(h_X,h_Z)$ plane. 
As can be seen, the shape of the topological phase is symmetric under the exchange of $h_X \leftrightarrow h_Z$. This is due to the fact that the perturbation $H_{0,1}+H_{1,0}$ respects the ``charge-flux symmetry" present in $H_{\rm TC}$. The transition lines that mark the boundaries of this phase are found to be either first-order or second-order lines.

Let us start with a discussion of the (two symmetric) first-order (cyan) lines
that are directly connected to the weakly first-order transition points (cyan
diamonds) of the three-state clock model in a (positive) transverse field. Near
these points, $h^{\rm c}$ and $h^{\star}$ are found to be very close so that it
is challenging to clearly decide whether the transition is first- or
second-order. However, given the precision reached and the existence of some
well-identified first-order points (not shown in \fref{fig:Higgs}) away from the cyan diamonds, it seems reasonable to argue that the two lines are likely first order.

The situation is different for the other part of the topological phase whose boundaries are connected to the second-order points (black diamonds) of the three-state clock  model in a (negative) transverse field. Here, the pCUT+iPEPS approach is always clearly consistent with a second-order phase transition.

In this region, the intersection of the two second-order transition lines
(green triangle)
is reminiscent of the phase diagram of the 
perturbed $\mathbb{Z}_2$ toric code in a  parallel magnetic field  where a
multi-critical point was found \cite{Vidal09_1,Tupitsyn10}. In the latter case,
this crossing point is also connected to a finite-length first-order line that
lies outside the topological phase. For the present problem ($N=3$), we have not
performed a small-$J$ perturbation theory likely to reveal a similar
feature. Nevertheless, we computed the gap exponent $z \,\nu$ along the
second-order (black) lines. As can be seen in \fref{fig:Higgs} (right), the behavior of this exponent is rather flat (same values as for $h_X=0$ or $h_Z=0$) except at its extremities where it increases significantly. This may indicate a different universality class at the crossing points (green triangle and magenta squares) as also found in the $N=2$ problem.
Anyway, let us stress that there are some limitations of our approach for the current problem. First, the perturbative expansion at order 7 is not sufficient to determine the gap exponent $z \,\nu$ accurately. Second, the finite width in $h^{\rm c}_X$ where the gap exponent differs from the value at $h^{\rm c}_X=0$ is likely an artifact of the finite-order series. Indeed, we rather expect that all points except crossing points belong to the same universality class as the anti-ferromagnetic clock model in a transverse field but one cannot exclude other scenarios. Once again, to obtain more quantitative results, it would be very valuable to perform numerical simulations of this model by means of alternative methods.

%
\subsubsection{The case $h_X=h_Z=0$ }\label{sssec:selfdual}
\begin{figure}[ht]
\centering
  \includegraphics[width=12cm]{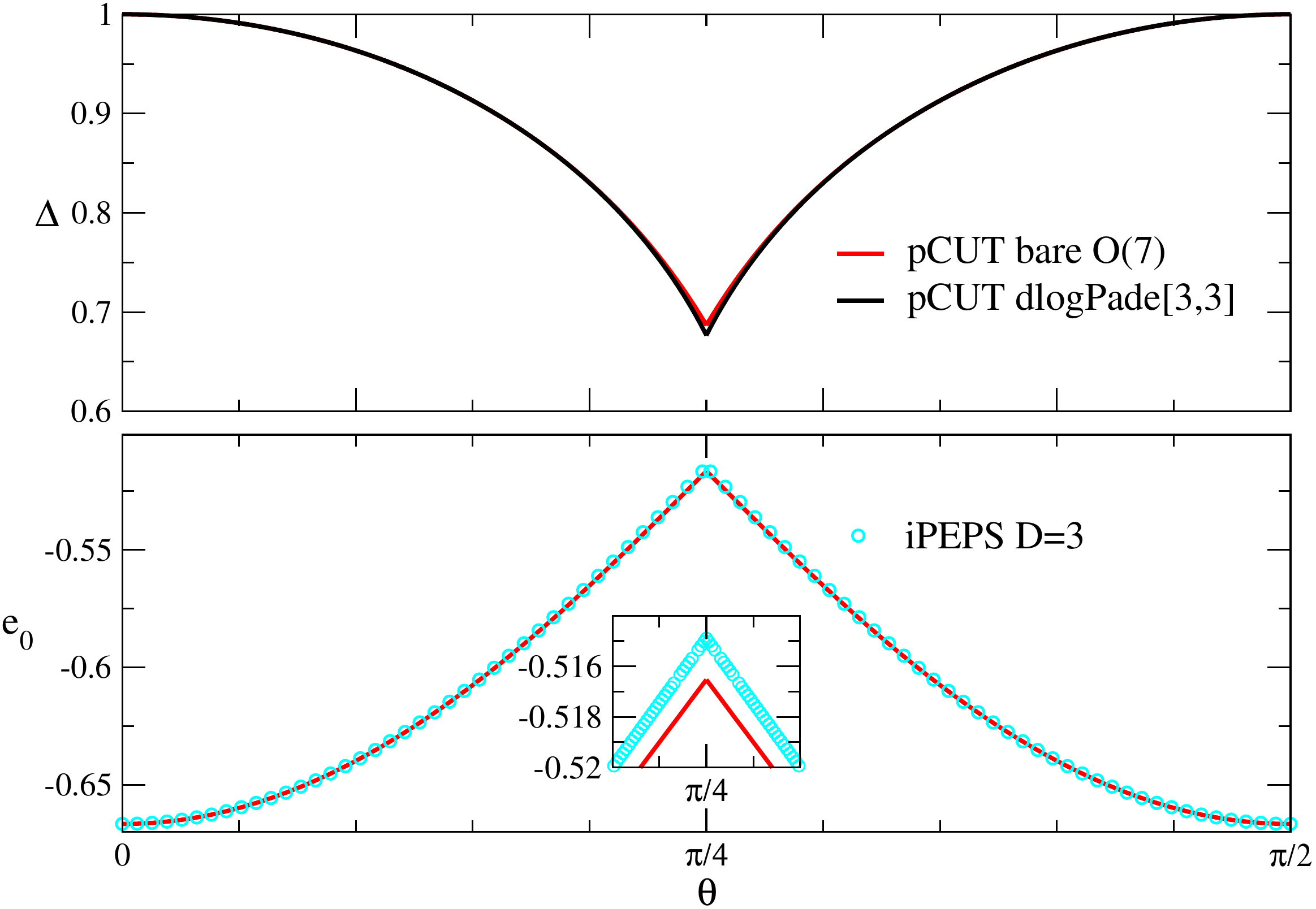} 
  \caption{Lower panel~: ground-state energy per site $e_0$ of $H(0,h_\perp,0)$ as a function of $\theta$ (with $J=\frac{1}{3}\cos\theta$ and $h_\perp=\frac{1}{3}\sin\theta$). Inset~: zoom of the kink at $\theta=\pi/4$ indicating the presence of a first-order phase transition. Upper panel~: 1-QP gap $\Delta$ as a function of  $\theta$ computed by pCUT. At  order 7,  other dlog-Pad\'{e} approximants of $\Delta$ are very close to the one shown here.}
  \label{fig:multi_flavor}
\end{figure}

%
 To conclude this analysis, let us consider the case of a ``transverse" perturbation that, as discussed in \sref{subsec:self} for general $N$,  leads to a self-dual spectrum for $H(0 , h_\perp,0)$. Since $J>0$ and $N$ is odd, one must also have $h_{\perp}>0$. A convenient parametrization for this problem consists in setting  $J=\frac{1}{3}\cos\theta$ and $h_{\perp}=\frac{1}{3}\sin\theta$ with $\theta \in [0,\pi/2]$. 
The unperturbed toric code  corresponds to $\theta=0$ whereas for $\theta=\pi/2$ the Hamiltonian is purely local. The self-dual point is located at $\theta=\pi/4$ ($J=h_{\perp}$) so that the spectrum is symmetric under the transformation $\theta \leftrightarrow \pi/2-\theta$.
Thanks to self-duality, one can determine directly the nature of the transition by studying the singularity of the ground-state energy. Note that we could not use this criterion in previous cases since we did not have reliable informations outside the topological phase. As can be clearly seen in \fref{fig:multi_flavor} (lower panel), the ground-state energy displays a kink at the self-dual point  ($\frac{\partial e_0}{\partial \theta}|_{\theta=\pi/4}$ is discontinuous) so that a first-order transition occurs there.

As explained in \sref{subsec:self}, the parity conservation rules constrain
the dynamics. In particular, they prevent a single QP to move in the presence of such a perturbation. Consequently the
1-QP energy level of the toric code ($\theta=0$) does not give rise to 
dispersive bands but is simply renormalized when $h_\perp \neq 0$. The
corresponding gap $\Delta$ is shown in   \fref{fig:multi_flavor} (upper panel)
and does not vanish at the self-dual point. However, this observation is compatible with the
existence of a first-order transition that is due to level crossings which cannot be captured by analyzing low-energy levels.
This situation is exactly the same as the one discussed in \cite{Vidal09_2} for the $\mathbb{Z}_2$ toric code in a transverse field. As in \cite{Vidal09_2}, 2-QP bound states are either pinned or mobile in one dimension only, while the simplest two-dimensional dispersing object is a 4-QP bound state.

\section{Conclusion and perspectives}\label{sec:Conclusion}
In this work, we have introduced exactly solvable models with
$\mathbb{Z}_N$ (Abelian) anyons that generalize Kitaev's famous toric code \cite{Kitaev03} or, equivalently, Wen's plaquette model \cite{Wen03}.
Our main motivation was to probe the robustness of  $\mathbb{Z}_{N>2}$
topologically-ordered phases following  recent works on the case $N=2$
\cite{Trebst07,Hamma08,Tupitsyn10,Vidal09_1,Vidal09_2,Dusuel11}. This question
is crucial since such phases are believed to be protected against external
perturbations up to an order proportional to the typical system size. Besides, this robustness has been proven recently for any local perturbation \cite{Klich10,Bravyi10}. However, as early noticed by Kitaev in his seminal paper \cite{Kitaev03}~: ``Of course, the perturbation should be small enough, or else a phase transition may occur."

To investigate the limits of the standard aforementioned perturbative argument, we added local perturbations to this $\mathbb{Z}_N$ toric code and we obtained several important results. First of all, for specific choices of the perturbation, the perturbed 
$\mathbb{Z}_N$ Hamiltonian can be mapped onto the two-dimensional $N$-state quantum clock model in a transverse field. This mapping generalizes the correspondence between the $\mathbb{Z}_2$ toric code in a parallel field and the transverse-field Ising model \cite{Trebst07,Hamma08} to arbitrary $N$.  For $N=3$ and antiferromagnetic couplings we found evidence for a second-order phase transition but we have not been able to determine accurately its universality class. Let us simply note that our estimate of the critical exponent $z\, \nu$ is compatible with that of the three-dimensional classical $XY$ model describing the three-dimensional three-state classical antiferromagnetic Potts model (see, {\it e.g.}, \cite{Gottlob94}).
Second, we have also shown that for $N=3$, multi-critical points are likely present in the phase diagram of $H(h_X,0,h_Z)$ [defined in (\ref{Eq:Hamiltonian_Z3TCF})] although, here again, it would be important to analyze them with complementary tools. In particular, numerical simulations would be as valuable as for $N=2$ \cite{Tupitsyn10}.
In addition, as already observed for the $N=2$ case \cite{Vidal09_2}, we
have shown that self-dual Hamiltonians may arise provided the perturbating
operators satisfy some constraints that we derived for arbitrary $N$. In this
case, the self-duality is associated with a dimensional reduction and, for $N=3$, we found that a first-order transition occurs at the self-dual point (as was also the case for $N=2$).

From a methodological point of view, the combination of high-order perturbation theory (pCUT) and variational techniques (iPEPS) has been shown to be very efficient in a domain where few alternative approaches exist. From that respect, let us mention that improvement of the variational ansatz could certainly be achieved by taking into account the gauge symmetry in the tensor network as recently discussed in \cite{Chen10,Tagliacozzo11,Swingle10}.

To conclude, we would like to mention related problems that would be worth
investigating in order to deepen our understanding of topological phases' fragility. In the simplest case $N=2$, it would already be worthwhile considering a perturbed toric code on a lattice which is not self-dual such as, for instance, the honeycomb lattice. Indeed, in this case, the role of charges and fluxes degrees of freedom cannot be exchanged by a simple lattice transformation and it is likely that such a model could lead to a non-trivial phase diagram in the presence of a uniform magnetic field.
One may also directly perturb  Wen's plaquette model with arbitrary local perturbations. Indeed, as shown in 
\cite{Kou08} for $N=2$, although Wen's and Kitaev's models are (almost) the same in the absence of perturbation, they display very different properties in the presence of a uniform magnetic field.
To go beyond $N=2$, we believe that the study of the breakdown of
$\mathbb{Z}_N$ topological phases initiated here would greatly benefit from a
large-$N$ analysis that might be performed in a field-theoretical framework.
Although it is clearly beyond the scope of this paper it may shed light on
several issues as for instance the universality class that
can be met at the boundaries of topological phases. Let us mention that recent studies suggested the possibility that conformal quantum critical points may describe these phase transitions (see for instance \cite{Ardonne04,Isakov11}). Finally, the most important step certainly consists in analyzing the robustness of non-Abelian topological phases \cite{Gils09,Burnell12} which are of direct interest for topological quantum computation but, undoubtedly, these quantum objects are much more tricky to handle.

\ack

We would like to thank K. Coester for fruitful discussions and for his help in the graph decompositions and M. Kamfor for a careful reading of the manuscript.
KPS acknowledges ESF and EuroHorcs for funding through his EURYI. 
RO acknowledges EU for funding through his Marie Curie IIF.

\appendix
\section{Series expansion in the  limit  $|h_Z|\ll J $ for $h_X=h_\perp=0$}\label{ssec:Details_Potts}

Setting $J=1/3$, the series expansion at order 11 of the ground-state energy per site of $H(0,0,h_Z)$ reads
\begin{eqnarray}\label{eq:e0_potts}
	\fl e_0&=&-\frac{2}{3}-2\, h_Z^2-h_Z^3-\frac{17}{2}\, h_Z^4-\frac{847}{36}\, h_Z^5-\frac{18407}{144}\, h_Z^6-\frac{15290}{27}\, h_Z^7 -\frac{995278817}{311040}\, h_Z^8 \nonumber\\
	\fl && -\frac{19860676421}{1166400}\, h_Z^9 -\frac{113666455562393}{1119744000}\, h_Z^{10}-\frac{9981578694811559}{16796160000}\, h_Z^{11}.
\end{eqnarray}
\\

The series expansion at order 11 of the 1-QP gap reads
\begin{eqnarray}\label{eq:delta+_potts}
	\fl \Delta&=&
    1-4\, \left| h_{{Z}} \right| -4\,h_{{Z}} \left| h_{{Z}} \right| -6\,{h_{{Z}}}^{2}-5\,{h_{{Z}}}^{2} \left| h_{{Z}} \right| -155\,{h_{{Z}}}^{3} \left| h_{{Z}} \right| -{\frac {965}{6}}\,{h_{{Z}}}^{4}\nonumber\\
    \fl && +{\frac {6469}{18}}\,{h_{{Z}}}^{4} \left| h_{{Z}} \right| +{\frac {1273}{3}}\,{h_{{Z}}}^{5}-{\frac {222718}{27}}\,{h_{{Z}}}^{5} \left| h_{{Z}} \right| -{\frac {597005}{72}}\,{h_{{Z}}}^{6}\nonumber\\
    \fl && +{\frac {17397799}{432}}\,{h_{{Z}}}^{6} \left| h_{{Z}} \right| +{\frac {54071671}{1296}}\,{h_{{Z}}}^{7}-{\frac {8607425885}{15552}}\,{h_{{Z}}}^{7} \left| h_{{Z}} \right| -{\frac {595589309}{1080}}\,{h_{{Z}}}^{8}\nonumber\\
    \fl && +{\frac {956588586169}{259200}}\,{h_{{Z}}}^{8} \left| h_{{Z}} \right| +{\frac {8712501685859}{2332800}}\,{h_{{Z}}}^{9}-{\frac {736434737832197}{17496000}}\,{h_{{Z}}}^{9} \left| h_{{Z}} \right|\nonumber\\
    \fl &&  -{\frac {46950055731802549}{1119744000}}\,{h_{{Z}}}^{10}+{\frac {277183060218266393}{839808000}}\,{h_{{Z}}}^{10} \left| h_{{Z}} \right|\nonumber\\
    \fl &&  +{\frac {61908400265234507}{186624000}}\,{h_{{Z}}}^{11}.
\end{eqnarray}
\\

\begin{table}[htp]
\caption{\label{tab:dlogpotts}
Table of critical values and gap exponents  obtained from dlog-Pad\'e
$[n,m]$ extrapolation of the 1-QP gap $\Delta$. Defective approximants,
\textit{i.e.}, the ones with spurious poles between zero and $h^{\rm c}_Z$, are marked with an asterisk. For $h_Z^c>0$ we do not give the corresponding exponent since, in this case,  a first-order transition occurs at $h^{\star}<h_Z^c$.}
\begin{indented} \item[]
\begin{tabular}{@{} l l l l | l l l l}
\br
$\left[n,m\right]$ & $h^{\rm c}_Z$ & $h^{\rm c}_Z$ & $z\nu$ & $\left[n,m\right]$ & $h^{\rm c}_Z$ & $h^{\rm c}_Z$ & $z\nu$\\ \mr
$\left[1,2\right]$ & 0.1322 &  -0.1982 & 0.791 & $\left[1,8\right]$ & 0.1286 &  -0.1947 & 0.716 \\ \cline{5 - 8}
$\left[2,3\right]$ & 0.1296 &  -0.1989 & 0.796 & $\left[1,9\right]$ & 0.1285 &  -0.1947 & 0.715 \\ \cline{5 - 8}
$\left[3,4\right]$ & 0.1296*&  -0.1970 & 0.771 & $\left[2,2\right]$ & 0.1284 &  -0.1996 & 0.806 \\
$\left[4,5\right]$ & 0.1287 &  -0.1944 & 0.706 & $\left[3,3\right]$ & 0.1295 &  -0.2053*& 0.802*\\ \cline{1 - 4}
$\left[1,3\right]$ & 0.1286 &  -0.1996 & 0.806 & $\left[4,4\right]$ & 0.1240 &  -0.1943 & 0.702 \\
$\left[2,4\right]$ & 0.1295 &  -0.2071*& 0.786*& $\left[5,5\right]$ & 0.1283 &  -0.1942*& 0.700*\\ \cline{5 - 8}
$\left[3,5\right]$ & 0.1262 &  -0.1945 & 0.708 & $\left[3,2\right]$ & 0.1301 &  -0.1977 & 0.776 \\
$\left[4,6\right]$ & 0.1284 &  -0.1945 & 0.707 & $\left[4,3\right]$ & 0.1290 &  -0.1951 & 0.726 \\ \cline{1 - 4}
$\left[1,4\right]$ & 0.1299 &  -0.1976 & 0.776 & $\left[5,4\right]$ & 0.1283 &  -0.1944 & 0.705 \\ \cline{5 - 8}
$\left[2,5\right]$ & 0.1288 &  -0.1953 & 0.731 & $\left[4,2\right]$ & 0.1292 &  -0.1957 & 0.739 \\
$\left[3,6\right]$ & 0.1282 &  -0.1944 & 0.706 & $\left[5,3\right]$ & 0.1293 &  -0.1970*& 0.748*\\ \cline{1 - 4}
$\left[1,5\right]$ & 0.1294 &  -0.1958 & 0.743 & $\left[6,4\right]$ & 0.1283 &  -0.1946 & 0.713 \\ \cline{5 - 8}
$\left[2,6\right]$ & 0.1213 &  -0.1966*& 0.751*& $\left[5,2\right]$ & 0.1291 &  -0.1953 & 0.729 \\
$\left[3,7\right]$ & 0.1285 &  -0.1947 & 0.715 & $\left[6,3\right]$ & 0.1286 &  -0.1946 & 0.712 \\ \cline{1 - 4} \cline{5 - 8}
$\left[1,6\right]$ & 0.1291 &  -0.1954 & 0.734 & $\left[6,2\right]$ & 0.1287 &  -0.1947 & 0.714 \\
$\left[2,7\right]$ & 0.1285 &  -0.1947 & 0.715 & $\left[7,3\right]$ & 0.1289*&  -0.1946 & 0.713 \\ \cline{1 - 4}\cline{5 - 8}
$\left[1,7\right]$ & 0.1288 &  -0.1949 & 0.720 & $\left[7,2\right]$ & 0.1286 &  -0.1946 & 0.713 \\ \cline{5 - 8}
$\left[2,8\right]$ & 0.1279 &  -0.1947 & 0.715 & $\left[8,2\right]$ & 0.1285 &  -0.1946 & 0.713 \\ \br
\end{tabular}
\end{indented}
\end{table}

\section{Series expansion in the  limit  $|h_X|,|h_Z|\ll J $ for $h_\perp=0$}\label{ssec:Details_Higgs}

Setting $J=1/3$, the series expansion at order 7 of the ground-state energy per site of $H(h_X,0,h_Z)$ reads
\begin{eqnarray}\label{eq:e0_higgs}
	\fl e_0 &=& -\frac{2}{3}-2\,S_{{2}}-S_{{3}}-\frac{17}{2}\,S_{{4}}+\frac{3}{2}\,P_{{4}}-{\frac {847}{36}}\,S_{{5}}+{\frac {21}{16}}\,P_{{4}}S_{{1}}-{\frac {18407}{144}}\,S_{{6}}\nonumber\\
    \fl && +{\frac {33}{32}}\,P_{{6}}+{\frac {933}{40}}\,P_{{4}}S_{{2}}-{\frac {15290}{27}}\,S_{{7}}+{\frac {1477451}{19200}}\,P_{{4}}S_{{3}}+{\frac {72039}{3200}}\,P_{{6}}S_{{1}}\, ,
\end{eqnarray}
with $S_n =h_X^n+h_Z^n$ and $P_n=h_X^{n/2}h_Z^{n/2}$.

For $N=3$ and for nonvanishing $h_X$ and $h_Z$, the 1-QP gap is defined as $\Delta=\min(\Delta_c,\Delta_f)$ where $\Delta_c$ ($\Delta_f$) denotes the charge (flux) gap.  
The series expansion at order 7 for the charge gap reads
\begin{eqnarray}\label{eq:delta-_higgs}
	\fl &\Delta_c&=
	1-4   \left| h_Z \right| -4  h_Z  \left| h_Z \right| -6  {h_Z}^{2}-5     \left| h_Z \right|  ^{3}+3  {h_X}^{2} \left| h_Z \right|\nonumber\\ \fl && 
	-155  h_Z    \left| h_Z \right|  ^{3}-{\frac {965}{6}}  {h_Z}^{4}+5  {h_X}^{2}h_Z  \left| h_Z \right| +7  {h_X}^{2}{h_Z}^{2}+3  {h_X}^{3} \left| h_Z \right| \nonumber\\ \fl && 
	+\frac{15}{2}  {h_X}^{4}+{\frac {6469}{18}}     \left| h_Z \right|  ^{5}+{\frac {1273}{3}}  {h_Z}^{5}+{\frac {85}{6}}  {h_X}^{2}   \left| h_Z \right|  ^{3}-{\frac {109}{24}}  {h_X}^{2}{h_Z}^{3}\nonumber\\ \fl && 
	+\frac{13}{3}  {h_X}^{3}h_Z  \left| h_Z \right| +{\frac {151}{24}}  {h_X}^{3}{h_Z}^{2}+{\frac {107}{4}}  {h_X}^{4} \left| h_Z \right| +{\frac {71}{3}}  {h_X}^{5}\nonumber\\ \fl && 
	-{\frac {222718}{27}}  h_Z    \left| h_Z \right|  ^{5}-{\frac {597005}{72}}  {h_Z}^{6}+{\frac {337873}{720}}  {h_X}^{2}h_Z    \left| h_Z \right|  ^{3}\nonumber\\ \fl && 
	+{\frac {676969}{1440}}  {h_X}^{2}{h_Z}^{4}+{\frac {521}{36}}  {h_X}^{3}   \left| h_Z \right|  ^{3}-{\frac {1043}{144}}  {h_X}^{3}{h_Z}^{3}+{\frac {2471}{36}}  {h_X}^{4}h_Z \left| h_Z \right| \nonumber\\ \fl && 
	+{\frac {12887}{90}}  {h_X}^{4}{h_Z}^{2}+{\frac {811}{9}}  {h_X}^{5} \left| h_Z \right| +{\frac {9373}{48}}  {h_X}^{6}+{\frac {17397799}{432}}     \left| h_Z \right|  ^{7}\nonumber\\ \fl && 
	+{\frac {54071671}{1296}}  {h_Z}^{7}-{\frac {6698749}{4320}}  {h_X}^{2}   \left| h_Z \right|  ^{5}-{\frac {160743271}{86400}}  {h_X}^{2}{h_Z}^{5}\nonumber\\ \fl && 
	+{\frac {2281303}{4800}}  {h_X}^{3}h_Z     \left| h_Z \right|  ^{3}+{\frac {8944457}{19200}}  {h_X}^{3}{h_Z}^{4}+{\frac {7637683}{43200}}  {h_X}^{4}   \left| h_Z \right|  ^{3}\nonumber\\ \fl && 
	+{\frac {415247}{14400}}  {h_X}^{4}{h_Z}^{3}+{\frac {20635}{108}}  {h_X}^{5}h_Z \left| h_Z \right| +{\frac {41043187}{86400}}  {h_X}^{5}{h_Z}^{2}\nonumber\\ \fl && 
	+{\frac {1191533}{1728}}  {h_X}^{6} \left| h_Z \right| +{\frac {8947}{9}}  {h_X}^{7}  .
\end{eqnarray}
The flux gap $\Delta_f$ is simply obtained from $\Delta_c$ by exchanging $h_X$ and $h_Z$.

\section{Series expansion in the  limit  $0<h_{\perp}\ll J $ for $h_X=h_Z=0$}\label{ssec:Details_selfdual}

Setting $J=\frac{1}{3}\cos\theta$, $h_{\perp}=\frac{1}{3}\sin \theta$, and denoting $t=\tan \theta=h_\perp/J\ll 1$, the series expansion at order~7 of the ground-state energy per site of $H(0,h_\perp,0)$ reads
\begin{eqnarray}\label{eq:e0_selfdual}
	\fl \quad \frac{e_0}{\cos \theta }&=&  
	-\frac{2}{3}
	-\frac{1}{18}\,  t^{2}
	-{\frac {1}{216}}\,  t^{3} 
	-{\frac {599}{272160}}\,  t^{4}
	-{\frac {209}{259200}}\,   t^{5}      
	-{\frac {896417}{2204496000}}\,   t^{6} \nonumber\\ \fl&&
	-{\frac {7184765443}{33331979520000}}\, t^{7}  \, .
\end{eqnarray}

The  series expansion at order 7 of the 1-QP gap reads
\begin{eqnarray}\label{eq:gap_selfdual}
	\fl \quad \frac{\Delta}{\cos   \theta  }&=& 
	1
	-{\frac {4}{27}}\,   t^{2} 
	-{\frac {5}{162}}\,   t^{3}
	-{\frac {617}{34992}}\,   t^{4}
	-{\frac {26773}{2624400}}\,   t^{5}       
	-{\frac {24667735793}{3888730944000}}\,   t^{6}  \nonumber\\ \fl&&
	-{\frac {2975252249029}{699971569920000}}\,   t^{7}    \, .
\end{eqnarray}

Since this model is self-dual, the corresponding quantities in the opposite limit $h_{\perp}\gg J>0 $ are obtained by changing $\theta$ into $\frac{\pi}{2}-\theta$.

\FloatBarrier

\def\newblock{\hskip .11em plus .33em minus .07em} 


\begin{thebibliography}{63}%
\makeatletter
\providecommand \@ifxundefined [1]{%
 \@ifx{#1\undefined}
}%
\providecommand \@ifnum [1]{%
 \ifnum #1\expandafter \@firstoftwo
 \else \expandafter \@secondoftwo
 \fi
}%
\providecommand \@ifx [1]{%
 \ifx #1\expandafter \@firstoftwo
 \else \expandafter \@secondoftwo
 \fi
}%
\providecommand \natexlab [1]{#1}%
\providecommand \enquote  [1]{``#1''}%
\providecommand \bibnamefont  [1]{#1}%
\providecommand \bibfnamefont [1]{#1}%
\providecommand \citenamefont [1]{#1}%
\providecommand \href@noop [0]{\@secondoftwo}%
\providecommand \href [0]{\begingroup \@sanitize@url \@href}%
\providecommand \@href[1]{\@@startlink{#1}\@@href}%
\providecommand \@@href[1]{\endgroup#1\@@endlink}%
\providecommand \@sanitize@url [0]{\catcode `\\12\catcode `\$12\catcode
  `\&12\catcode `\#12\catcode `\^12\catcode `\_12\catcode `\%12\relax}%
\providecommand \@@startlink[1]{}%
\providecommand \@@endlink[0]{}%
\providecommand \url  [0]{\begingroup\@sanitize@url \@url }%
\providecommand \@url [1]{\endgroup\@href {#1}{\urlprefix }}%
\providecommand \urlprefix  [0]{URL }%
\providecommand \Eprint [0]{\href }%
\providecommand \doibase [0]{http://dx.doi.org/}%
\providecommand \selectlanguage [0]{\@gobble}%
\providecommand \bibinfo  [0]{\@secondoftwo}%
\providecommand \bibfield  [0]{\@secondoftwo}%
\providecommand \translation [1]{[#1]}%
\providecommand \BibitemOpen [0]{}%
\providecommand \bibitemStop [0]{}%
\providecommand \bibitemNoStop [0]{.\EOS\space}%
\providecommand \EOS [0]{\spacefactor3000\relax}%
\providecommand \BibitemShut  [1]{\csname bibitem#1\endcsname}%
\let\auto@bib@innerbib\@empty
\bibitem [{\citenamefont {Wen}(1989)}]{Wen89}%
  \BibitemOpen
  \bibfield  {author} {\bibinfo {author} {\bibfnamefont {X.-G.}\ \bibnamefont
  {Wen}},\ }\href {\doibase 10.1103/PhysRevB.40.7387} {\bibfield  {journal}
  {\bibinfo  {journal} {Phys. Rev. B}\ }\textbf {\bibinfo {volume} {40}},\
  \bibinfo {pages} {7387} (\bibinfo {year} {1989})}\BibitemShut {NoStop}%
\bibitem [{\citenamefont {Wen}(1990)}]{Wen90_1}%
  \BibitemOpen
  \bibfield  {author} {\bibinfo {author} {\bibfnamefont {X.-G.}\ \bibnamefont
  {Wen}},\ }\href {\doibase 10.1142/S0217979290000139} {\bibfield  {journal}
  {\bibinfo  {journal} {Int. J. Mod. Phys. B}\ }\textbf {\bibinfo {volume}
  {4}},\ \bibinfo {pages} {239} (\bibinfo {year} {1990})}\BibitemShut {NoStop}%
\bibitem [{\citenamefont {Kitaev}(2003)}]{Kitaev03}%
  \BibitemOpen
  \bibfield  {author} {\bibinfo {author} {\bibfnamefont {A.~Y.}\ \bibnamefont
  {Kitaev}},\ }\href {\doibase 10.1016/S0003-4916(02)00018-0} {\bibfield
  {journal} {\bibinfo  {journal} {Ann. Phys. (N.Y.)}\ }\textbf {\bibinfo
  {volume} {303}},\ \bibinfo {pages} {2} (\bibinfo {year} {2003})}\BibitemShut
  {NoStop}%
\bibitem [{\citenamefont {Klich}(2010)}]{Klich10}%
  \BibitemOpen
  \bibfield  {author} {\bibinfo {author} {\bibfnamefont {I.}~\bibnamefont
  {Klich}},\ }\href {\doibase 10.1016/j.aop.2010.05.002} {\bibfield  {journal}
  {\bibinfo  {journal} {Ann. Phys. (N.Y.)}\ }\textbf {\bibinfo {volume}
  {325}},\ \bibinfo {pages} {120} (\bibinfo {year} {2010})}\BibitemShut
  {NoStop}%
\bibitem [{\citenamefont {Bravyi}, \citenamefont {Hastings},\ and\
  \citenamefont {Michalakis}(2010)}]{Bravyi10}%
  \BibitemOpen
  \bibfield  {author} {\bibinfo {author} {\bibfnamefont {S.}~\bibnamefont
  {Bravyi}}, \bibinfo {author} {\bibfnamefont {M.~B.}\ \bibnamefont
  {Hastings}}, \ and\ \bibinfo {author} {\bibfnamefont {S.}~\bibnamefont
  {Michalakis}},\ }\href {\doibase 10.1063/1.3490195} {\bibfield  {journal}
  {\bibinfo  {journal} {J. Math. Phys.}\ }\textbf {\bibinfo {volume} {51}},\
  \bibinfo {pages} {093512} (\bibinfo {year} {2010})}\BibitemShut {NoStop}%
\bibitem [{\citenamefont {Ogburn}\ and\ \citenamefont
  {Preskill}(1999)}]{Ogburn99}%
  \BibitemOpen
  \bibfield  {author} {\bibinfo {author} {\bibfnamefont {W.~R.}\ \bibnamefont
  {Ogburn}}\ and\ \bibinfo {author} {\bibfnamefont {J.}~\bibnamefont
  {Preskill}},\ }\href {\doibase 10.1007/3-540-49208-9_31} {\bibfield
  {journal} {\bibinfo  {journal} {Lect. Notes Comput. Sci.}\ }\textbf {\bibinfo
  {volume} {1509}},\ \bibinfo {pages} {341} (\bibinfo {year}
  {1999})}\BibitemShut {NoStop}%
\bibitem [{\citenamefont {Castelnovo}\ and\ \citenamefont
  {Chamon}(2007)}]{Castelnovo07}%
  \BibitemOpen
  \bibfield  {author} {\bibinfo {author} {\bibfnamefont {C.}~\bibnamefont
  {Castelnovo}}\ and\ \bibinfo {author} {\bibfnamefont {C.}~\bibnamefont
  {Chamon}},\ }\href {\doibase 10.1103/PhysRevB.76.184442} {\bibfield
  {journal} {\bibinfo  {journal} {Phys. Rev. B}\ }\textbf {\bibinfo {volume}
  {76}},\ \bibinfo {pages} {184442} (\bibinfo {year} {2007})}\BibitemShut
  {NoStop}%
\bibitem [{\citenamefont {Nussinov}\ and\ \citenamefont
  {Ortiz}(2008)}]{Nussinov08}%
  \BibitemOpen
  \bibfield  {author} {\bibinfo {author} {\bibfnamefont {Z.}~\bibnamefont
  {Nussinov}}\ and\ \bibinfo {author} {\bibfnamefont {G.}~\bibnamefont
  {Ortiz}},\ }\href {\doibase 10.1103/PhysRevB.77.064302} {\bibfield  {journal}
  {\bibinfo  {journal} {Phys. Rev. B}\ }\textbf {\bibinfo {volume} {77}},\
  \bibinfo {pages} {064302} (\bibinfo {year} {2008})}\BibitemShut {NoStop}%
\bibitem [{\citenamefont {Alicki}, \citenamefont {Fannes},\ and\ \citenamefont
  {Horodecki}(2009)}]{Alicki09}%
  \BibitemOpen
  \bibfield  {author} {\bibinfo {author} {\bibfnamefont {R.}~\bibnamefont
  {Alicki}}, \bibinfo {author} {\bibfnamefont {M.}~\bibnamefont {Fannes}}, \
  and\ \bibinfo {author} {\bibfnamefont {M.}~\bibnamefont {Horodecki}},\ }\href
  {\doibase 10.1088/1751-8113/42/6/065303} {\bibfield  {journal} {\bibinfo
  {journal} {J. Phys. A}\ }\textbf {\bibinfo {volume} {42}},\ \bibinfo {pages}
  {065303} (\bibinfo {year} {2009})}\BibitemShut {NoStop}%
\bibitem [{\citenamefont {Iblisdir}\ \emph {et~al.}(2009)\citenamefont
  {Iblisdir}, \citenamefont {P\'erez-Garc\'ia}, \citenamefont {Aguado},\ and\
  \citenamefont {Pachos}}]{Iblisdir09}%
  \BibitemOpen
  \bibfield  {author} {\bibinfo {author} {\bibfnamefont {S.}~\bibnamefont
  {Iblisdir}}, \bibinfo {author} {\bibfnamefont {D.}~\bibnamefont
  {P\'erez-Garc\'ia}}, \bibinfo {author} {\bibfnamefont {M.}~\bibnamefont
  {Aguado}}, \ and\ \bibinfo {author} {\bibfnamefont {J.}~\bibnamefont
  {Pachos}},\ }\href {\doibase 10.1103/PhysRevB.79.134303} {\bibfield
  {journal} {\bibinfo  {journal} {Phys. Rev. B}\ }\textbf {\bibinfo {volume}
  {79}},\ \bibinfo {pages} {134303} (\bibinfo {year} {2009})}\BibitemShut
  {NoStop}%
\bibitem [{\citenamefont {Iblisdir}\ \emph {et~al.}(2010)\citenamefont
  {Iblisdir}, \citenamefont {P\'erez-Garc\'ia}, \citenamefont {Aguado},\ and\
  \citenamefont {Pachos}}]{Iblisdir10}%
  \BibitemOpen
  \bibfield  {author} {\bibinfo {author} {\bibfnamefont {S.}~\bibnamefont
  {Iblisdir}}, \bibinfo {author} {\bibfnamefont {D.}~\bibnamefont
  {P\'erez-Garc\'ia}}, \bibinfo {author} {\bibfnamefont {M.}~\bibnamefont
  {Aguado}}, \ and\ \bibinfo {author} {\bibfnamefont {J.}~\bibnamefont
  {Pachos}},\ }\href {\doibase 10.1016/j.nuclphysb.2009.11.009} {\bibfield
  {journal} {\bibinfo  {journal} {Nucl. Phys. B}\ }\textbf {\bibinfo {volume}
  {829}},\ \bibinfo {pages} {401} (\bibinfo {year} {2010})}\BibitemShut
  {NoStop}%
\bibitem [{\citenamefont {Wootton}\ and\ \citenamefont
  {Pachos}(2011)}]{Wootton11}%
  \BibitemOpen
  \bibfield  {author} {\bibinfo {author} {\bibfnamefont {J.~R.}\ \bibnamefont
  {Wootton}}\ and\ \bibinfo {author} {\bibfnamefont {J.~K.}\ \bibnamefont
  {Pachos}},\ }\href {\doibase 10.1103/PhysRevLett.107.030503} {\bibfield
  {journal} {\bibinfo  {journal} {Phys. Rev. Lett.}\ }\textbf {\bibinfo
  {volume} {107}},\ \bibinfo {pages} {030503} (\bibinfo {year}
  {2011})}\BibitemShut {NoStop}%
\bibitem [{\citenamefont {Stark}\ \emph {et~al.}(2011)\citenamefont {Stark},
  \citenamefont {Pollet}, \citenamefont {Imamo\u{g}lu},\ and\ \citenamefont
  {Renner}}]{Stark11}%
  \BibitemOpen
  \bibfield  {author} {\bibinfo {author} {\bibfnamefont {C.}~\bibnamefont
  {Stark}}, \bibinfo {author} {\bibfnamefont {L.}~\bibnamefont {Pollet}},
  \bibinfo {author} {\bibfnamefont {A.}~\bibnamefont {Imamo\u{g}lu}}, \ and\
  \bibinfo {author} {\bibfnamefont {R.}~\bibnamefont {Renner}},\ }\href
  {\doibase 10.1103/PhysRevLett.107.030504} {\bibfield  {journal} {\bibinfo
  {journal} {Phys. Rev. Lett.}\ }\textbf {\bibinfo {volume} {107}},\ \bibinfo
  {pages} {030504} (\bibinfo {year} {2011})}\BibitemShut {NoStop}%
\bibitem [{\citenamefont {Trebst}\ \emph {et~al.}(2007)\citenamefont {Trebst},
  \citenamefont {Werner}, \citenamefont {Troyer}, \citenamefont {Shtengel},\
  and\ \citenamefont {Nayak}}]{Trebst07}%
  \BibitemOpen
  \bibfield  {author} {\bibinfo {author} {\bibfnamefont {S.}~\bibnamefont
  {Trebst}}, \bibinfo {author} {\bibfnamefont {P.}~\bibnamefont {Werner}},
  \bibinfo {author} {\bibfnamefont {M.}~\bibnamefont {Troyer}}, \bibinfo
  {author} {\bibfnamefont {K.}~\bibnamefont {Shtengel}}, \ and\ \bibinfo
  {author} {\bibfnamefont {C.}~\bibnamefont {Nayak}},\ }\href {\doibase
  10.1103/PhysRevLett.98.070602} {\bibfield  {journal} {\bibinfo  {journal}
  {Phys. Rev. Lett.}\ }\textbf {\bibinfo {volume} {98}},\ \bibinfo {pages}
  {070602} (\bibinfo {year} {2007})}\BibitemShut {NoStop}%
\bibitem [{\citenamefont {Hamma}\ and\ \citenamefont {Lidar}(2008)}]{Hamma08}%
  \BibitemOpen
  \bibfield  {author} {\bibinfo {author} {\bibfnamefont {A.}~\bibnamefont
  {Hamma}}\ and\ \bibinfo {author} {\bibfnamefont {D.~A.}\ \bibnamefont
  {Lidar}},\ }\href {\doibase 10.1103/PhysRevLett.100.030502} {\bibfield
  {journal} {\bibinfo  {journal} {Phys. Rev. Lett.}\ }\textbf {\bibinfo
  {volume} {100}},\ \bibinfo {pages} {030502} (\bibinfo {year}
  {2008})}\BibitemShut {NoStop}%
\bibitem [{\citenamefont {Tupitsyn}\ \emph {et~al.}(2010)\citenamefont
  {Tupitsyn}, \citenamefont {Kitaev}, \citenamefont {Prokof'ev},\ and\
  \citenamefont {Stamp}}]{Tupitsyn10}%
  \BibitemOpen
  \bibfield  {author} {\bibinfo {author} {\bibfnamefont {I.~S.}\ \bibnamefont
  {Tupitsyn}}, \bibinfo {author} {\bibfnamefont {A.}~\bibnamefont {Kitaev}},
  \bibinfo {author} {\bibfnamefont {N.~V.}\ \bibnamefont {Prokof'ev}}, \ and\
  \bibinfo {author} {\bibfnamefont {P.~C.~E.}\ \bibnamefont {Stamp}},\ }\href
  {\doibase 10.1103/PhysRevB.82.085114} {\bibfield  {journal} {\bibinfo
  {journal} {Phys. Rev. B}\ }\textbf {\bibinfo {volume} {82}},\ \bibinfo
  {pages} {085114} (\bibinfo {year} {2010})}\BibitemShut {NoStop}%
\bibitem [{\citenamefont {Vidal}, \citenamefont {Dusuel},\ and\ \citenamefont
  {Schmidt}(2009)}]{Vidal09_1}%
  \BibitemOpen
  \bibfield  {author} {\bibinfo {author} {\bibfnamefont {J.}~\bibnamefont
  {Vidal}}, \bibinfo {author} {\bibfnamefont {S.}~\bibnamefont {Dusuel}}, \
  and\ \bibinfo {author} {\bibfnamefont {K.~P.}\ \bibnamefont {Schmidt}},\
  }\href {\doibase 10.1103/PhysRevB.79.033109} {\bibfield  {journal} {\bibinfo
  {journal} {Phys. Rev. B}\ }\textbf {\bibinfo {volume} {79}},\ \bibinfo
  {pages} {033109} (\bibinfo {year} {2009})}\BibitemShut {NoStop}%
\bibitem [{\citenamefont {Vidal}\ \emph {et~al.}(2009)\citenamefont {Vidal},
  \citenamefont {Thomale}, \citenamefont {Schmidt},\ and\ \citenamefont
  {Dusuel}}]{Vidal09_2}%
  \BibitemOpen
  \bibfield  {author} {\bibinfo {author} {\bibfnamefont {J.}~\bibnamefont
  {Vidal}}, \bibinfo {author} {\bibfnamefont {R.}~\bibnamefont {Thomale}},
  \bibinfo {author} {\bibfnamefont {K.~P.}\ \bibnamefont {Schmidt}}, \ and\
  \bibinfo {author} {\bibfnamefont {S.}~\bibnamefont {Dusuel}},\ }\href
  {\doibase 10.1103/PhysRevB.80.081104} {\bibfield  {journal} {\bibinfo
  {journal} {Phys. Rev. B}\ }\textbf {\bibinfo {volume} {80}},\ \bibinfo
  {pages} {081104(R)} (\bibinfo {year} {2009})}\BibitemShut {NoStop}%
\bibitem [{\citenamefont {Dusuel}\ \emph {et~al.}(2011)\citenamefont {Dusuel},
  \citenamefont {Kamfor}, \citenamefont {Or\'us}, \citenamefont {Schmidt},\
  and\ \citenamefont {Vidal}}]{Dusuel11}%
  \BibitemOpen
  \bibfield  {author} {\bibinfo {author} {\bibfnamefont {S.}~\bibnamefont
  {Dusuel}}, \bibinfo {author} {\bibfnamefont {M.}~\bibnamefont {Kamfor}},
  \bibinfo {author} {\bibfnamefont {R.}~\bibnamefont {Or\'us}}, \bibinfo
  {author} {\bibfnamefont {K.~P.}\ \bibnamefont {Schmidt}}, \ and\ \bibinfo
  {author} {\bibfnamefont {J.}~\bibnamefont {Vidal}},\ }\href {\doibase
  10.1103/PhysRevLett.106.107203} {\bibfield  {journal} {\bibinfo  {journal}
  {Phys. Rev. Lett.}\ }\textbf {\bibinfo {volume} {106}},\ \bibinfo {pages}
  {107203} (\bibinfo {year} {2011})}\BibitemShut {NoStop}%
\bibitem [{\citenamefont {Tagliacozzo}\ and\ \citenamefont
  {Vidal}(2011)}]{Tagliacozzo11}%
  \BibitemOpen
  \bibfield  {author} {\bibinfo {author} {\bibfnamefont {L.}~\bibnamefont
  {Tagliacozzo}}\ and\ \bibinfo {author} {\bibfnamefont {G.}~\bibnamefont
  {Vidal}},\ }\href {\doibase 10.1103/PhysRevB.83.115127} {\bibfield  {journal}
  {\bibinfo  {journal} {Phys. Rev. B}\ }\textbf {\bibinfo {volume} {83}},\
  \bibinfo {pages} {115127} (\bibinfo {year} {2011})}\BibitemShut {NoStop}%
\bibitem [{\citenamefont {Levin}\ and\ \citenamefont {Wen}(2005)}]{Levin05}%
  \BibitemOpen
  \bibfield  {author} {\bibinfo {author} {\bibfnamefont {M.~A.}\ \bibnamefont
  {Levin}}\ and\ \bibinfo {author} {\bibfnamefont {X.-G.}\ \bibnamefont
  {Wen}},\ }\href {\doibase 10.1103/PhysRevB.71.045110} {\bibfield  {journal}
  {\bibinfo  {journal} {Phys. Rev. B}\ }\textbf {\bibinfo {volume} {71}},\
  \bibinfo {pages} {045110} (\bibinfo {year} {2005})}\BibitemShut {NoStop}%
\bibitem [{\citenamefont {Burnell}, \citenamefont {Simon},\ and\ \citenamefont
  {Slingerland}(2012)}]{Burnell12}%
  \BibitemOpen
  \bibfield  {author} {\bibinfo {author} {\bibfnamefont {F.~J.}\ \bibnamefont
  {Burnell}}, \bibinfo {author} {\bibfnamefont {S.~H.}\ \bibnamefont {Simon}},
  \ and\ \bibinfo {author} {\bibfnamefont {J.~K.}\ \bibnamefont
  {Slingerland}},\ }\href {\doibase 10.1088/1367-2630/14/1/015004} {\bibfield
  {journal} {\bibinfo  {journal} {New J. Phys.}\ }\textbf {\bibinfo {volume}
  {14}},\ \bibinfo {pages} {015004} (\bibinfo {year} {2012})}\BibitemShut
  {NoStop}%
\bibitem [{\citenamefont {Burnell}, \citenamefont {Simon},\ and\ \citenamefont
  {Slingerland}(2011)}]{Burnell11}%
  \BibitemOpen
  \bibfield  {author} {\bibinfo {author} {\bibfnamefont {F.~J.}\ \bibnamefont
  {Burnell}}, \bibinfo {author} {\bibfnamefont {S.~H.}\ \bibnamefont {Simon}},
  \ and\ \bibinfo {author} {\bibfnamefont {J.~K.}\ \bibnamefont
  {Slingerland}},\ }\href {\doibase 10.1103/PhysRevB.84.125434} {\bibfield
  {journal} {\bibinfo  {journal} {Phys. Rev. B}\ }\textbf {\bibinfo {volume}
  {84}},\ \bibinfo {pages} {125434} (\bibinfo {year} {2011})}\BibitemShut
  {NoStop}%
\bibitem [{\citenamefont {Bais}\ and\ \citenamefont
  {Slingerland}(2009)}]{Bais09}%
  \BibitemOpen
  \bibfield  {author} {\bibinfo {author} {\bibfnamefont {F.~A.}\ \bibnamefont
  {Bais}}\ and\ \bibinfo {author} {\bibfnamefont {J.~K.}\ \bibnamefont
  {Slingerland}},\ }\href {\doibase 10.1103/PhysRevB.79.045316} {\bibfield
  {journal} {\bibinfo  {journal} {Phys. Rev. B}\ }\textbf {\bibinfo {volume}
  {79}},\ \bibinfo {pages} {045316} (\bibinfo {year} {2009})}\BibitemShut
  {NoStop}%
\bibitem [{\citenamefont {Wootton}\ \emph {et~al.}(2011)\citenamefont
  {Wootton}, \citenamefont {Lahtinen}, \citenamefont {Dou\c{c}ot},\ and\
  \citenamefont {Pachos}}]{Wootton11_1}%
  \BibitemOpen
  \bibfield  {author} {\bibinfo {author} {\bibfnamefont {J.~R.}\ \bibnamefont
  {Wootton}}, \bibinfo {author} {\bibfnamefont {V.}~\bibnamefont {Lahtinen}},
  \bibinfo {author} {\bibfnamefont {B.}~\bibnamefont {Dou\c{c}ot}}, \ and\
  \bibinfo {author} {\bibfnamefont {J.~K.}\ \bibnamefont {Pachos}},\ }\href
  {\doibase 10.1016/j.aop.2011.05.008} {\bibfield  {journal} {\bibinfo
  {journal} {Ann. Phys. (N.Y.)}\ }\textbf {\bibinfo {volume} {326}},\ \bibinfo
  {pages} {2307} (\bibinfo {year} {2011})}\BibitemShut {NoStop}%
\bibitem [{\citenamefont {Bullock}\ and\ \citenamefont
  {Brennen}(2007)}]{Bullock07}%
  \BibitemOpen
  \bibfield  {author} {\bibinfo {author} {\bibfnamefont {S.~S.}\ \bibnamefont
  {Bullock}}\ and\ \bibinfo {author} {\bibfnamefont {G.~K.}\ \bibnamefont
  {Brennen}},\ }\href {\doibase 10.1088/1751-8113/40/13/013} {\bibfield
  {journal} {\bibinfo  {journal} {J. Phys. A}\ }\textbf {\bibinfo {volume}
  {40}},\ \bibinfo {pages} {3481} (\bibinfo {year} {2007})}\BibitemShut
  {NoStop}%
\bibitem [{\citenamefont {Wen}(2003)}]{Wen03}%
  \BibitemOpen
  \bibfield  {author} {\bibinfo {author} {\bibfnamefont {X.-G.}\ \bibnamefont
  {Wen}},\ }\href {\doibase 10.1103/PhysRevLett.90.016803} {\bibfield
  {journal} {\bibinfo  {journal} {Phys. Rev. Lett.}\ }\textbf {\bibinfo
  {volume} {90}},\ \bibinfo {pages} {016803} (\bibinfo {year}
  {2003})}\BibitemShut {NoStop}%
\bibitem [{\citenamefont {Hamer}, \citenamefont {Oitmaa},\ and\ \citenamefont
  {Weihong}(1992)}]{Hamer92}%
  \BibitemOpen
  \bibfield  {author} {\bibinfo {author} {\bibfnamefont {C.~J.}\ \bibnamefont
  {Hamer}}, \bibinfo {author} {\bibfnamefont {J.}~\bibnamefont {Oitmaa}}, \
  and\ \bibinfo {author} {\bibfnamefont {Z.}~\bibnamefont {Weihong}},\ }\href
  {\doibase 10.1088/0305-4470/25/7/023} {\bibfield  {journal} {\bibinfo
  {journal} {J. Phys. A}\ }\textbf {\bibinfo {volume} {25}},\ \bibinfo {pages}
  {1821} (\bibinfo {year} {1992})}\BibitemShut {NoStop}%
\bibitem [{\citenamefont {Kitaev}(2006)}]{Kitaev06}%
  \BibitemOpen
  \bibfield  {author} {\bibinfo {author} {\bibfnamefont {A.}~\bibnamefont
  {Kitaev}},\ }\href {\doibase 10.1016/j.aop.2005.10.005} {\bibfield  {journal}
  {\bibinfo  {journal} {Ann. Phys. (N.Y.)}\ }\textbf {\bibinfo {volume}
  {321}},\ \bibinfo {pages} {2} (\bibinfo {year} {2006})}\BibitemShut {NoStop}%
\bibitem [{\citenamefont {Vidal}, \citenamefont {Schmidt},\ and\ \citenamefont
  {Dusuel}(2008)}]{Vidal08_2}%
  \BibitemOpen
  \bibfield  {author} {\bibinfo {author} {\bibfnamefont {J.}~\bibnamefont
  {Vidal}}, \bibinfo {author} {\bibfnamefont {K.~P.}\ \bibnamefont {Schmidt}},
  \ and\ \bibinfo {author} {\bibfnamefont {S.}~\bibnamefont {Dusuel}},\ }\href
  {\doibase 10.1103/PhysRevB.78.245121} {\bibfield  {journal} {\bibinfo
  {journal} {Phys. Rev. B}\ }\textbf {\bibinfo {volume} {78}},\ \bibinfo
  {pages} {245121} (\bibinfo {year} {2008})}\BibitemShut {NoStop}%
\bibitem [{\citenamefont {Arakawa}\ and\ \citenamefont
  {Ichinose}(2004)}]{Arakawa04}%
  \BibitemOpen
  \bibfield  {author} {\bibinfo {author} {\bibfnamefont {G.}~\bibnamefont
  {Arakawa}}\ and\ \bibinfo {author} {\bibfnamefont {I.}~\bibnamefont
  {Ichinose}},\ }\href {\doibase 10} {\bibfield  {journal} {\bibinfo  {journal}
  {Ann. Phys. (N.Y.)}\ }\textbf {\bibinfo {volume} {311}},\ \bibinfo {pages}
  {152} (\bibinfo {year} {2004})}\BibitemShut {NoStop}%
\bibitem [{\citenamefont {Dou\c{c}ot}, \citenamefont {Ioffe},\ and\
  \citenamefont {Vidal}(2004)}]{Doucot04}%
  \BibitemOpen
  \bibfield  {author} {\bibinfo {author} {\bibfnamefont {B.}~\bibnamefont
  {Dou\c{c}ot}}, \bibinfo {author} {\bibfnamefont {L.~B.}\ \bibnamefont
  {Ioffe}}, \ and\ \bibinfo {author} {\bibfnamefont {J.}~\bibnamefont
  {Vidal}},\ }\href {\doibase 10.1103/PhysRevB.69.214501} {\bibfield  {journal}
  {\bibinfo  {journal} {Phys. Rev. B}\ }\textbf {\bibinfo {volume} {69}},\
  \bibinfo {pages} {214501} (\bibinfo {year} {2004})}\BibitemShut {NoStop}%
\bibitem [{\citenamefont {Cobanera}, \citenamefont {Ortiz},\ and\ \citenamefont
  {Nussinov}(2011)}]{Cobanera11}%
  \BibitemOpen
  \bibfield  {author} {\bibinfo {author} {\bibfnamefont {E.}~\bibnamefont
  {Cobanera}}, \bibinfo {author} {\bibfnamefont {G.}~\bibnamefont {Ortiz}}, \
  and\ \bibinfo {author} {\bibfnamefont {Z.}~\bibnamefont {Nussinov}},\ }\href
  {\doibase 10.1080/00018732.2011.619814} {\bibfield  {journal} {\bibinfo
  {journal} {Adv. Phys.}\ }\textbf {\bibinfo {volume} {60}},\ \bibinfo {pages}
  {679} (\bibinfo {year} {2011})}\BibitemShut {NoStop}%
\bibitem [{\citenamefont {Chen}\ and\ \citenamefont {Hu}(2007)}]{Chen07}%
  \BibitemOpen
  \bibfield  {author} {\bibinfo {author} {\bibfnamefont {H.-D.}\ \bibnamefont
  {Chen}}\ and\ \bibinfo {author} {\bibfnamefont {J.}~\bibnamefont {Hu}},\
  }\href {\doibase 10.1103/PhysRevB.76.193101} {\bibfield  {journal} {\bibinfo
  {journal} {Phys. Rev. B}\ }\textbf {\bibinfo {volume} {76}},\ \bibinfo
  {pages} {193101} (\bibinfo {year} {2007})}\BibitemShut {NoStop}%
\bibitem [{\citenamefont {Xu}\ and\ \citenamefont {Moore}(2004)}]{Xu04}%
  \BibitemOpen
  \bibfield  {author} {\bibinfo {author} {\bibfnamefont {C.}~\bibnamefont
  {Xu}}\ and\ \bibinfo {author} {\bibfnamefont {J.~E.}\ \bibnamefont {Moore}},\
  }\href {\doibase 10.1103/PhysRevLett.93.047003} {\bibfield  {journal}
  {\bibinfo  {journal} {Phys. Rev. Lett.}\ }\textbf {\bibinfo {volume} {93}},\
  \bibinfo {pages} {047003} (\bibinfo {year} {2004})}\BibitemShut {NoStop}%
\bibitem [{\citenamefont {Xu}\ and\ \citenamefont {Moore}(2005)}]{Xu05}%
  \BibitemOpen
  \bibfield  {author} {\bibinfo {author} {\bibfnamefont {C.}~\bibnamefont
  {Xu}}\ and\ \bibinfo {author} {\bibfnamefont {J.~E.}\ \bibnamefont {Moore}},\
  }\href {\doibase 10.1016/j.nuclphysb.2005.04.003} {\bibfield  {journal}
  {\bibinfo  {journal} {Nucl. Phys. B}\ }\textbf {\bibinfo {volume} {716}},\
  \bibinfo {pages} {487} (\bibinfo {year} {2005})}\BibitemShut {NoStop}%
\bibitem [{\citenamefont {Nussinov}\ and\ \citenamefont
  {Fradkin}(2005)}]{Nussinov05}%
  \BibitemOpen
  \bibfield  {author} {\bibinfo {author} {\bibfnamefont {Z.}~\bibnamefont
  {Nussinov}}\ and\ \bibinfo {author} {\bibfnamefont {E.}~\bibnamefont
  {Fradkin}},\ }\href {\doibase 10.1103/PhysRevB.71.195120} {\bibfield
  {journal} {\bibinfo  {journal} {Phys. Rev. B}\ }\textbf {\bibinfo {volume}
  {71}},\ \bibinfo {pages} {195120} (\bibinfo {year} {2005})}\BibitemShut
  {NoStop}%
\bibitem [{\citenamefont {Nussinov}, \citenamefont {Ortiz},\ and\ \citenamefont
  {Cobanera}()}]{Nussinov11}%
  \BibitemOpen
  \bibfield  {author} {\bibinfo {author} {\bibfnamefont {Z.}~\bibnamefont
  {Nussinov}}, \bibinfo {author} {\bibfnamefont {G.}~\bibnamefont {Ortiz}}, \
  and\ \bibinfo {author} {\bibfnamefont {E.}~\bibnamefont {Cobanera}},\
  }\href@noop {} {}\bibinfo {note}
  {\href{http://arxiv.org/abs/1110.2179}{arXiv:1110.2179}}\BibitemShut
  {NoStop}%
\bibitem [{\citenamefont {Wegner}(1994)}]{Wegner94}%
  \BibitemOpen
  \bibfield  {author} {\bibinfo {author} {\bibfnamefont {F.}~\bibnamefont
  {Wegner}},\ }\href {\doibase 10.1002/andp.19945060203} {\bibfield  {journal}
  {\bibinfo  {journal} {Ann. Phys. (Leipzig)}\ }\textbf {\bibinfo {volume}
  {3}},\ \bibinfo {pages} {77} (\bibinfo {year} {1994})}\BibitemShut {NoStop}%
\bibitem [{\citenamefont {Knetter}\ and\ \citenamefont
  {Uhrig}(2000)}]{Knetter00}%
  \BibitemOpen
  \bibfield  {author} {\bibinfo {author} {\bibfnamefont {C.}~\bibnamefont
  {Knetter}}\ and\ \bibinfo {author} {\bibfnamefont {G.~S.}\ \bibnamefont
  {Uhrig}},\ }\href {\doibase 10.1007/s100510050026} {\bibfield  {journal}
  {\bibinfo  {journal} {Eur. Phys. J. B}\ }\textbf {\bibinfo {volume} {13}},\
  \bibinfo {pages} {209} (\bibinfo {year} {2000})}\BibitemShut {NoStop}%
\bibitem [{\citenamefont {Knetter}, \citenamefont {Schmidt},\ and\
  \citenamefont {Uhrig}(2003)}]{Knetter03}%
  \BibitemOpen
  \bibfield  {author} {\bibinfo {author} {\bibfnamefont {C.}~\bibnamefont
  {Knetter}}, \bibinfo {author} {\bibfnamefont {K.~P.}\ \bibnamefont
  {Schmidt}}, \ and\ \bibinfo {author} {\bibfnamefont {G.~S.}\ \bibnamefont
  {Uhrig}},\ }\href {\doibase 10.1088/0305-4470/36/29/302} {\bibfield
  {journal} {\bibinfo  {journal} {J. Phys. A}\ }\textbf {\bibinfo {volume}
  {36}},\ \bibinfo {pages} {7889} (\bibinfo {year} {2003})}\BibitemShut
  {NoStop}%
\bibitem [{\citenamefont {Dusuel}\ \emph {et~al.}(2010)\citenamefont {Dusuel},
  \citenamefont {Kamfor}, \citenamefont {Schmidt}, \citenamefont {Thomale},\
  and\ \citenamefont {Vidal}}]{Dusuel10}%
  \BibitemOpen
  \bibfield  {author} {\bibinfo {author} {\bibfnamefont {S.}~\bibnamefont
  {Dusuel}}, \bibinfo {author} {\bibfnamefont {M.}~\bibnamefont {Kamfor}},
  \bibinfo {author} {\bibfnamefont {K.~P.}\ \bibnamefont {Schmidt}}, \bibinfo
  {author} {\bibfnamefont {R.}~\bibnamefont {Thomale}}, \ and\ \bibinfo
  {author} {\bibfnamefont {J.}~\bibnamefont {Vidal}},\ }\href {\doibase
  10.1103/PhysRevB.81.064412} {\bibfield  {journal} {\bibinfo  {journal} {Phys.
  Rev. B}\ }\textbf {\bibinfo {volume} {81}},\ \bibinfo {pages} {064412}
  (\bibinfo {year} {2010})}\BibitemShut {NoStop}%
\bibitem [{\citenamefont {Oitmaa}, \citenamefont {Hamer},\ and\ \citenamefont
  {Zheng}(2006)}]{Oitmaa06}%
  \BibitemOpen
  \bibfield  {author} {\bibinfo {author} {\bibfnamefont {J.}~\bibnamefont
  {Oitmaa}}, \bibinfo {author} {\bibfnamefont {C.~J.}\ \bibnamefont {Hamer}}, \
  and\ \bibinfo {author} {\bibfnamefont {W.}~\bibnamefont {Zheng}},\
  }\href@noop {} {\emph {\bibinfo {title} {Series expansion methods for
  strongly interacting lattice models}}}\ (\bibinfo  {publisher} {Cambridge
  University Press},\ \bibinfo {year} {2006})\BibitemShut {NoStop}%
\bibitem [{\citenamefont {Jordan}\ \emph {et~al.}(2008)\citenamefont {Jordan},
  \citenamefont {Or\'us}, \citenamefont {Vidal}, \citenamefont {Verstraete},\
  and\ \citenamefont {Cirac}}]{Jordan08}%
  \BibitemOpen
  \bibfield  {author} {\bibinfo {author} {\bibfnamefont {J.}~\bibnamefont
  {Jordan}}, \bibinfo {author} {\bibfnamefont {R.}~\bibnamefont {Or\'us}},
  \bibinfo {author} {\bibfnamefont {G.}~\bibnamefont {Vidal}}, \bibinfo
  {author} {\bibfnamefont {F.}~\bibnamefont {Verstraete}}, \ and\ \bibinfo
  {author} {\bibfnamefont {J.~I.}\ \bibnamefont {Cirac}},\ }\href {\doibase
  10.1103/PhysRevLett.101.250602} {\bibfield  {journal} {\bibinfo  {journal}
  {Phys. Rev. Lett.}\ }\textbf {\bibinfo {volume} {101}},\ \bibinfo {pages}
  {250602} (\bibinfo {year} {2008})}\BibitemShut {NoStop}%
\bibitem [{\citenamefont {Corboz}\ \emph {et~al.}(2010)\citenamefont {Corboz},
  \citenamefont {Or\'us}, \citenamefont {Bauer},\ and\ \citenamefont
  {Vidal}}]{Corboz10}%
  \BibitemOpen
  \bibfield  {author} {\bibinfo {author} {\bibfnamefont {P.}~\bibnamefont
  {Corboz}}, \bibinfo {author} {\bibfnamefont {R.}~\bibnamefont {Or\'us}},
  \bibinfo {author} {\bibfnamefont {B.}~\bibnamefont {Bauer}}, \ and\ \bibinfo
  {author} {\bibfnamefont {G.}~\bibnamefont {Vidal}},\ }\href {\doibase
  10.1103/PhysRevB.81.165104} {\bibfield  {journal} {\bibinfo  {journal} {Phys.
  Rev. B}\ }\textbf {\bibinfo {volume} {81}},\ \bibinfo {pages} {165104}
  (\bibinfo {year} {2010})}\BibitemShut {NoStop}%
\bibitem [{\citenamefont {Jordan}()}]{Jordan11}%
  \BibitemOpen
  \bibfield  {author} {\bibinfo {author} {\bibfnamefont {J.}~\bibnamefont
  {Jordan}},\ }\href@noop {} {}\bibinfo {note} {Ph. D. thesis, The University
  of Queensland, Brisbane, Australia (2011), available at~:
  \href{http://www.romanorus.com/JordanThesis.pdf}{
  http://www.romanorus.com/JordanThesis.pdf}}\BibitemShut {NoStop}%
\bibitem [{\citenamefont {Or\'us}\ and\ \citenamefont
  {Vidal}(2009)}]{Orus09_2}%
  \BibitemOpen
  \bibfield  {author} {\bibinfo {author} {\bibfnamefont {R.}~\bibnamefont
  {Or\'us}}\ and\ \bibinfo {author} {\bibfnamefont {G.}~\bibnamefont {Vidal}},\
  }\href {\doibase 10.1103/PhysRevB.80.09443} {\bibfield  {journal} {\bibinfo
  {journal} {Phys. Rev. B}\ }\textbf {\bibinfo {volume} {80}},\ \bibinfo
  {pages} {094403} (\bibinfo {year} {2009})}\BibitemShut {NoStop}%
\bibitem [{\citenamefont {Corboz}\ \emph {et~al.}(2011)\citenamefont {Corboz},
  \citenamefont {White}, \citenamefont {Vidal},\ and\ \citenamefont
  {Troyer}}]{Corboz11}%
  \BibitemOpen
  \bibfield  {author} {\bibinfo {author} {\bibfnamefont {P.}~\bibnamefont
  {Corboz}}, \bibinfo {author} {\bibfnamefont {S.~R.}\ \bibnamefont {White}},
  \bibinfo {author} {\bibfnamefont {G.}~\bibnamefont {Vidal}}, \ and\ \bibinfo
  {author} {\bibfnamefont {M.}~\bibnamefont {Troyer}},\ }\href {\doibase
  10.1103/PhysRevB.84.041108} {\bibfield  {journal} {\bibinfo  {journal} {Phys.
  Rev. B}\ }\textbf {\bibinfo {volume} {84}},\ \bibinfo {pages} {041108}
  (\bibinfo {year} {2011})}\BibitemShut {NoStop}%
\bibitem [{\citenamefont {Verstraete}\ and\ \citenamefont
  {Cirac}()}]{Verstraete04}%
  \BibitemOpen
  \bibfield  {author} {\bibinfo {author} {\bibfnamefont {F.}~\bibnamefont
  {Verstraete}}\ and\ \bibinfo {author} {\bibfnamefont {J.~I.}\ \bibnamefont
  {Cirac}},\ }\href@noop {} {}\bibinfo {note}
  {\href{http://arxiv.org/abs/cond-mat/0407066}{arXiv:0407066}}\BibitemShut
  {NoStop}%
\bibitem [{\citenamefont {Verstraete}\ \emph {et~al.}(2006)\citenamefont
  {Verstraete}, \citenamefont {Wolf}, \citenamefont {Perez-Garcia},\ and\
  \citenamefont {Cirac}}]{Verstraete06}%
  \BibitemOpen
  \bibfield  {author} {\bibinfo {author} {\bibfnamefont {F.}~\bibnamefont
  {Verstraete}}, \bibinfo {author} {\bibfnamefont {M.~M.}\ \bibnamefont
  {Wolf}}, \bibinfo {author} {\bibfnamefont {D.}~\bibnamefont {Perez-Garcia}},
  \ and\ \bibinfo {author} {\bibfnamefont {J.~I.}\ \bibnamefont {Cirac}},\
  }\href {\doibase 10.1103/PhysRevLett.96.220601} {\bibfield  {journal}
  {\bibinfo  {journal} {Phys. Rev. Lett.}\ }\textbf {\bibinfo {volume} {96}},\
  \bibinfo {pages} {220601} (\bibinfo {year} {2006})}\BibitemShut {NoStop}%
\bibitem [{\citenamefont {Gu}, \citenamefont {Levin},\ and\ \citenamefont
  {Wen}(2008)}]{Gu08}%
  \BibitemOpen
  \bibfield  {author} {\bibinfo {author} {\bibfnamefont {Z.-C.}\ \bibnamefont
  {Gu}}, \bibinfo {author} {\bibfnamefont {M.}~\bibnamefont {Levin}}, \ and\
  \bibinfo {author} {\bibfnamefont {X.-G.}\ \bibnamefont {Wen}},\ }\href
  {\doibase 10.1103/PhysRevB.78.205116} {\bibfield  {journal} {\bibinfo
  {journal} {Phys. Rev. B}\ }\textbf {\bibinfo {volume} {78}},\ \bibinfo
  {pages} {205116} (\bibinfo {year} {2008})}\BibitemShut {NoStop}%
\bibitem [{\citenamefont {Yang}\ \emph {et~al.}(2010)\citenamefont {Yang},
  \citenamefont {L\"auchli}, \citenamefont {Mila},\ and\ \citenamefont
  {Schmidt}}]{Yang10}%
  \BibitemOpen
  \bibfield  {author} {\bibinfo {author} {\bibfnamefont {H.-Y.}\ \bibnamefont
  {Yang}}, \bibinfo {author} {\bibfnamefont {A.~M.}\ \bibnamefont {L\"auchli}},
  \bibinfo {author} {\bibfnamefont {F.}~\bibnamefont {Mila}}, \ and\ \bibinfo
  {author} {\bibfnamefont {K.~P.}\ \bibnamefont {Schmidt}},\ }\href {\doibase
  10.1103/PhysRevLett.105.267204} {\bibfield  {journal} {\bibinfo  {journal}
  {Phys. Rev. Lett.}\ }\textbf {\bibinfo {volume} {105}},\ \bibinfo {pages}
  {267204} (\bibinfo {year} {2010})}\BibitemShut {NoStop}%
\bibitem [{\citenamefont {Hamer}\ \emph {et~al.}(1990)\citenamefont {Hamer},
  \citenamefont {Aydin}, \citenamefont {Oitmaa},\ and\ \citenamefont
  {He}}]{Hamer90}%
  \BibitemOpen
  \bibfield  {author} {\bibinfo {author} {\bibfnamefont {C.~J.}\ \bibnamefont
  {Hamer}}, \bibinfo {author} {\bibfnamefont {M.}~\bibnamefont {Aydin}},
  \bibinfo {author} {\bibfnamefont {J.}~\bibnamefont {Oitmaa}}, \ and\ \bibinfo
  {author} {\bibfnamefont {H.-X.}\ \bibnamefont {He}},\ }\href {\doibase
  10.1088/0305-4470/23/17/031} {\bibfield  {journal} {\bibinfo  {journal} {J.
  Phys. A}\ }\textbf {\bibinfo {volume} {23}},\ \bibinfo {pages} {4025}
  (\bibinfo {year} {1990})}\BibitemShut {NoStop}%
\bibitem [{\citenamefont {Liu}\ \emph {et~al.}(2010)\citenamefont {Liu},
  \citenamefont {Wang}, \citenamefont {Sandvik}, \citenamefont {Su},\ and\
  \citenamefont {Kao}}]{Liu10}%
  \BibitemOpen
  \bibfield  {author} {\bibinfo {author} {\bibfnamefont {C.}~\bibnamefont
  {Liu}}, \bibinfo {author} {\bibfnamefont {L.}~\bibnamefont {Wang}}, \bibinfo
  {author} {\bibfnamefont {A.~W.}\ \bibnamefont {Sandvik}}, \bibinfo {author}
  {\bibfnamefont {Y.-C.}\ \bibnamefont {Su}}, \ and\ \bibinfo {author}
  {\bibfnamefont {Y.-J.}\ \bibnamefont {Kao}},\ }\href {\doibase
  10.1103/PhysRevB.82.060410} {\bibfield  {journal} {\bibinfo  {journal} {Phys.
  Rev. B}\ }\textbf {\bibinfo {volume} {82}},\ \bibinfo {pages} {060410}
  (\bibinfo {year} {2010})}\BibitemShut {NoStop}%
\bibitem [{\citenamefont {Dutta}\ \emph {et~al.}()\citenamefont {Dutta},
  \citenamefont {Divakaran}, \citenamefont {Sen}, \citenamefont {Chakrabarti},
  \citenamefont {Rosenbaum},\ and\ \citenamefont {Aeppli}}]{Dutta11}%
  \BibitemOpen
  \bibfield  {author} {\bibinfo {author} {\bibfnamefont {A.}~\bibnamefont
  {Dutta}}, \bibinfo {author} {\bibfnamefont {U.}~\bibnamefont {Divakaran}},
  \bibinfo {author} {\bibfnamefont {D.}~\bibnamefont {Sen}}, \bibinfo {author}
  {\bibfnamefont {B.~K.}\ \bibnamefont {Chakrabarti}}, \bibinfo {author}
  {\bibfnamefont {T.~F.}\ \bibnamefont {Rosenbaum}}, \ and\ \bibinfo {author}
  {\bibfnamefont {G.}~\bibnamefont {Aeppli}},\ }\href@noop {} {}\bibinfo {note}
  {\href{http://arxiv.org/abs/1012.0653}{arXiv:1012.0653}}\BibitemShut
  {NoStop}%
\bibitem [{\citenamefont {He}, \citenamefont {Hamer},\ and\ \citenamefont
  {Oitmaa}(1990)}]{He90}%
  \BibitemOpen
  \bibfield  {author} {\bibinfo {author} {\bibfnamefont {H.-X.}\ \bibnamefont
  {He}}, \bibinfo {author} {\bibfnamefont {C.~J.}\ \bibnamefont {Hamer}}, \
  and\ \bibinfo {author} {\bibfnamefont {J.}~\bibnamefont {Oitmaa}},\ }\href
  {\doibase 10.1088/0305-4470/23/10/018} {\bibfield  {journal} {\bibinfo
  {journal} {J. Phys. A}\ }\textbf {\bibinfo {volume} {23}},\ \bibinfo {pages}
  {1775} (\bibinfo {year} {1990})}\BibitemShut {NoStop}%
\bibitem [{\citenamefont {Gottlob}\ and\ \citenamefont
  {Hasenbusch}(1994)}]{Gottlob94}%
  \BibitemOpen
  \bibfield  {author} {\bibinfo {author} {\bibfnamefont {A.~P.}\ \bibnamefont
  {Gottlob}}\ and\ \bibinfo {author} {\bibfnamefont {M.}~\bibnamefont
  {Hasenbusch}},\ }\href {\doibase 10.1016/0378-4371(94)00097-2} {\bibfield
  {journal} {\bibinfo  {journal} {Physica A}\ }\textbf {\bibinfo {volume}
  {210}},\ \bibinfo {pages} {217} (\bibinfo {year} {1994})}\BibitemShut
  {NoStop}%
\bibitem [{\citenamefont {Chen}\ \emph {et~al.}(2010)\citenamefont {Chen},
  \citenamefont {Zeng}, \citenamefont {Gu}, \citenamefont {Chuang},\ and\
  \citenamefont {Wen}}]{Chen10}%
  \BibitemOpen
  \bibfield  {author} {\bibinfo {author} {\bibfnamefont {X.}~\bibnamefont
  {Chen}}, \bibinfo {author} {\bibfnamefont {B.}~\bibnamefont {Zeng}}, \bibinfo
  {author} {\bibfnamefont {Z.-C.}\ \bibnamefont {Gu}}, \bibinfo {author}
  {\bibfnamefont {I.~L.}\ \bibnamefont {Chuang}}, \ and\ \bibinfo {author}
  {\bibfnamefont {X.-G.}\ \bibnamefont {Wen}},\ }\href {\doibase
  10.1103/PhysRevB.82.165119} {\bibfield  {journal} {\bibinfo  {journal} {Phys.
  Rev. B}\ }\textbf {\bibinfo {volume} {82}},\ \bibinfo {pages} {165119}
  (\bibinfo {year} {2010})}\BibitemShut {NoStop}%
\bibitem [{\citenamefont {Swingle}\ and\ \citenamefont {Wen}()}]{Swingle10}%
  \BibitemOpen
  \bibfield  {author} {\bibinfo {author} {\bibfnamefont {B.}~\bibnamefont
  {Swingle}}\ and\ \bibinfo {author} {\bibfnamefont {X.-G.}\ \bibnamefont
  {Wen}},\ }\href@noop {} {}\bibinfo {note}
  {\href{http://arxiv.org/abs/1001.4517}{arXiv:1001.4517}}\BibitemShut
  {NoStop}%
\bibitem [{\citenamefont {Yu}, \citenamefont {Kou},\ and\ \citenamefont
  {Wen}(2008)}]{Kou08}%
  \BibitemOpen
  \bibfield  {author} {\bibinfo {author} {\bibfnamefont {J.}~\bibnamefont
  {Yu}}, \bibinfo {author} {\bibfnamefont {S.-P.}\ \bibnamefont {Kou}}, \ and\
  \bibinfo {author} {\bibfnamefont {X.-G.}\ \bibnamefont {Wen}},\ }\href
  {\doibase 10.1209/0295-5075/84/17004} {\bibfield  {journal} {\bibinfo
  {journal} {Europhys. Lett.}\ }\textbf {\bibinfo {volume} {84}},\ \bibinfo
  {pages} {17004} (\bibinfo {year} {2008})}\BibitemShut {NoStop}%
\bibitem [{\citenamefont {Ardonne}, \citenamefont {Fendley},\ and\
  \citenamefont {Fradkin}(2004)}]{Ardonne04}%
  \BibitemOpen
  \bibfield  {author} {\bibinfo {author} {\bibfnamefont {E.}~\bibnamefont
  {Ardonne}}, \bibinfo {author} {\bibfnamefont {P.}~\bibnamefont {Fendley}}, \
  and\ \bibinfo {author} {\bibfnamefont {E.}~\bibnamefont {Fradkin}},\
  }\href@noop {} {\bibfield  {journal} {\bibinfo  {journal} {Ann. Phys.
  (N.Y.)}\ }\textbf {\bibinfo {volume} {310}},\ \bibinfo {pages} {493}
  (\bibinfo {year} {2004})}\BibitemShut {NoStop}%
\bibitem [{\citenamefont {Isakov}\ \emph {et~al.}(2011)\citenamefont {Isakov},
  \citenamefont {Fendley}, \citenamefont {Ludwig}, \citenamefont {Trebst},\
  and\ \citenamefont {Troyer}}]{Isakov11}%
  \BibitemOpen
  \bibfield  {author} {\bibinfo {author} {\bibfnamefont {S.~V.}\ \bibnamefont
  {Isakov}}, \bibinfo {author} {\bibfnamefont {P.}~\bibnamefont {Fendley}},
  \bibinfo {author} {\bibfnamefont {A.~W.~W.}\ \bibnamefont {Ludwig}}, \bibinfo
  {author} {\bibfnamefont {S.}~\bibnamefont {Trebst}}, \ and\ \bibinfo {author}
  {\bibfnamefont {M.}~\bibnamefont {Troyer}},\ }\href {\doibase
  10.1103/PhysRevB.83.125114} {\bibfield  {journal} {\bibinfo  {journal} {Phys.
  Rev. B}\ }\textbf {\bibinfo {volume} {83}},\ \bibinfo {pages} {125114}
  (\bibinfo {year} {2011})}\BibitemShut {NoStop}%
\bibitem [{\citenamefont {C.Gils}\ \emph {et~al.}(2009)\citenamefont {C.Gils},
  \citenamefont {Trebst}, \citenamefont {Kitaev}, \citenamefont {Ludwig},
  \citenamefont {Troyer},\ and\ \citenamefont {Wang}}]{Gils09}%
  \BibitemOpen
  \bibfield  {author} {\bibinfo {author} {\bibnamefont {C.Gils}}, \bibinfo
  {author} {\bibfnamefont {S.}~\bibnamefont {Trebst}}, \bibinfo {author}
  {\bibfnamefont {A.}~\bibnamefont {Kitaev}}, \bibinfo {author} {\bibfnamefont
  {A.~W.~W.}\ \bibnamefont {Ludwig}}, \bibinfo {author} {\bibfnamefont
  {M.}~\bibnamefont {Troyer}}, \ and\ \bibinfo {author} {\bibfnamefont
  {Z.}~\bibnamefont {Wang}},\ }\href {\doibase 10.1038/nphys1396} {\bibfield
  {journal} {\bibinfo  {journal} {Nat. Phys.}\ }\textbf {\bibinfo {volume}
  {5}},\ \bibinfo {pages} {834} (\bibinfo {year} {2009})}\BibitemShut {NoStop}%
\end{thebibliography}

%

\end{document}